\documentclass[journal=jctcce,manuscript=article]{achemso}
\usepackage{graphicx}
\usepackage[table]{xcolor}
\usepackage{subcaption}
\usepackage[version=4]{mhchem} 
\usepackage[colorlinks=true,linkcolor=black,citecolor=black,urlcolor=black]{hyperref}
\usepackage{xr}
\usepackage{hyperref}
\usepackage{cleveref}
\setkeys{acs}{doi = true}
\usepackage{siunitx}
\DeclareSIUnit\angstrom{\text {Å}}

\renewcommand\vec{\mathbf}

\newcommand{\zenodo}{\href{http://doi.org/10.5281/zenodo.10570083}{10.5281/zenodo.10570083}}
\newcommand{\edgek}{$\mathrm{K}$}
\newcommand{\edgelone}{$\mathrm{L_1}$}

\author{Jessica A. Martinez B.}
\affiliation{Department of Chemistry, Rutgers University, Newark, New Jersey}

\author{Matteo De Santis}
\affiliation{Univ. Lille, CNRS, UMR 8523 - PhLAM - Physique des Lasers Atomes et Molécules, F-59000 Lille, France}

\author{Michele Pavanello}
\affiliation{Department of Physics, Rutgers University, Newark, New Jersey}   
\altaffiliation{Department of Chemistry, Rutgers University, Newark, New Jersey}

\author{Valérie Vallet}
\affiliation{Univ. Lille, CNRS, UMR 8523 - PhLAM - Physique des Lasers Atomes et Molécules, F-59000 Lille, France}

\author{André Severo Pereira Gomes}
\email{andre.gomes@univ-lille.fr}
\affiliation{Univ. Lille, CNRS, UMR 8523 - PhLAM - Physique des Lasers Atomes et Molécules, F-59000 Lille, France}

\title{Solvation effects on halides core spectra with Multilevel Real-Time quantum embedding}

\abbreviations{RT, TD-DFT, BOMME, FDE}
\keywords{RT, TD-DFT, BOMME, FDE}

\begin{document}

\begin{tocentry}
\includegraphics[width=8.4cm,height=4.2cm]{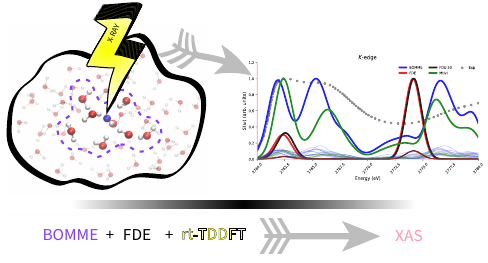}

\end{tocentry}

\begin{abstract}
Quantum embedding techniques are gaining popularity as they provide an accurate description of electronic systems at a reduced computational cost. In this work, we introduce a novel subsystem-based electronic structure embedding method that combines the projection-based block-orthogonalized Manby-Miller embedding (BOMME) with the density-based Frozen Density Embedding (FDE) methods. Our approach is especially effective for systems in which the building blocks (or subsystems) interact at varying strengths while still maintaining a lower computational cost compared to a quantum simulation of the entire system. To evaluate the performance of our method, we assess its ability to reproduce the X-ray absorption spectra (XAS) of chloride and fluoride anions in aqueous solutions (based on a \num{50}-water droplet model) via real-time time-dependent density functional theory (rt-TDDFT) calculations. We employ an ensemble approach to compute XAS for the {\edgek} and {\edgelone} edges, utilizing multiple snapshots of configuration space obtained from classical molecular dynamics simulations with a polarizable force field.  Configurational averaging influences both the broadening of spectral features and their intensities, with contributions to the final intensities originating from different geometry configurations. We found that embedding models that are too approximate for halide-water specific interactions, as in the case of FDE, fail to reproduce the experimental spectrum for chloride. Meanwhile, the BOMME approach tends to overestimate intensities, particularly for higher energy features because of finite-size effects.  Combining FDE for the second solvation shell and retaining BOMME for the first solvation shell mitigates this effect, resulting in an overall improved agreement within the energy range of the experimental spectrum. Additionally, we compute the transition densities of the relevant transitions, confirming that these transitions occur within the halide systems. Thus, our real-time QM/QM/QM embedding method proves to be a promising approach for modeling XAS of solvated systems.

\end{abstract}

\section{Introduction}
The aqueous dynamics of anions such as halides and other halogenated species have attracted the interest of many scientists eager to uncover their multiple electronic processes \cite{muralidharan_quasi-chemical_2019,hofer_solvation_2022,chowdhuri_dynamics_2006,heuft_density_2005,bakker_structural_2008}. These processes depend both on the intrinsic characteristics of the ions themselves and their interaction with the surrounding hydration shells~\cite{real_structural_2016,antalek_solvation_2016,marin_failure_2021,waluyo_different_2014}. Particularly noteworthy is the strong influence on the extended structure of water resulting from the short-range interactions between ions and their surrounding waters~\cite{antalek_solvation_2016,collins_behavior_2019}. Ion solvation effects underpin the development of innovative technologies such as rechargeable fluoride-ion batteries \cite{fang_long-life_2023} and capacitive deionization electrodes (CDI) \cite{mao_molecular_2022}. For the latter, it has been demonstrated that the local distribution of the hydrogen bond network created by the anion directly retards the dielectric response of the CDI electrode. 
	
Among the methods that can be used to characterize the behavior of solvated ions, those probing core electrons, such as X-ray absorption spectroscopy (XAS), are particularly well-suited to get an insight into the structural and electronic behavior of anionic systems, due to the great sensitivity of core states to even the slightest perturbations in the electronic structure of the absorbing center (such as those provoked by changes in their local environment)~\cite{antalek_solvation_2016,kulik_local_2010}. Interpreting XAS experiments is challenging due to the myriad physical processes at play, necessitating theoretical models to make sense of experimental observations. 

Among theoretical approaches, effective theories (e.g.\ crystal or ligand-field models, impurity models, etc) are widely used~\cite{alayoglu_uncovering_2023,zhang_autonomous_2023} as they efficiently account for environment, electron correlation, and relativistic effects, offering a cost-effective way to simulate X-ray absorption near edge structure (XANES) and long-range multiple scattering (EXAFS) regions~\cite{zhao_quantitative_2007,tofoni_p_2023}. Both techniques have been used to simulate the spectra of aqueous iodide, bromide, and chloride \cite{antalek_solvation_2016}. A alternative flexible method for simulating XANES is found in molecular electronic structure methods and among them those based on time-dependent density functional theory employing the restricted energy window formalism (REW-TDDFT\cite{herbert_time-dependent_2023}, referred to hereafter as TDDFT), due to their favorable balance between computational cost and accuracy. While TD-DFT is known to underestimate the absolute value of absorption energies due to the lack of core hole relaxation \cite{misael_core_2023, park_mixed-reference_2023}, experimental spectral features can be accurately reproduced once a global shift is applied\cite{fransson_calibrating_2023,liekhus-schmaltz_ultrafast_2021} to the features of the edge under consideration.

Standard TDDFT calculations, however, can be computationally expensive for large-scale systems involving a solute accompanied by tens to hundreds of solvent molecules, especially when temperature effects are taken into account via molecular dynamics calculations. In such cases, embedding approaches prove useful in reducing computational costs by allowing the tailoring of accuracy for different parts of the system (subsystems) based on their importance to the property of interest. 

The use of a classical representation for the solvent and solvent-solute interactions, as done in widely popular approaches such as QM/MM (quantum mechanics/molecular mechanics) model \cite{hofer_solvation_2022,pezeshki_molecular_2014} or the quasi-chemical theory (QCT) with polarizable continuum model (PCM) \cite{gomez_hydrated_2022}, brings about substantial computational savings. However, it may potentially miss effects due the quantum mechanical nature of the solvent and, in the case of PCM, struggle to capture directional interactions like hydrogen bonds~\cite{andreussi_continuum_2019}. These shortcomings can be alleviated by extending the QM-level region to include a number of solvent molecules, though this approach does not guarantee a quick convergence of calculated spectra with the number of explicit solvent molecules\cite{milanese_convergence_2017}.

	

Purely quantum embedding (QE) techniques~\cite{ding_embedded_2017,sharma_efficient_2022,wesolowski_frozen-density_2015,iannuzzi_density_2006} have emerged more recently, with the aim to retain a fully quantum mechanical description for all subsystems. In approaches such as the frozen density embedding (FDE)~\cite{krishtal_subsystem_2015,jacob_subsystem_2014,wesolowski_frozen-density_2015,fu_excitation_2023,niemeyer_subsystem_2023} method, the aim is to represent the interaction between subsystems by a (local) embedding potential constructed from subsystem electron densities. This allows for linear scaling calculations with respect to the number of subsystems\cite{sen_towards_2023,krishtal_subsystem_2015}, as well as for very straightforward combinations of electronic structure methods (wavefunction-based, DFT-based, etc) for the computation of valence~\cite{bouchafra_predictive_2018} and core~\cite{opoku_simulating_2022} ionizations and excitations~\cite{misael_core_2023,hofener_solvatochromic_2013}. However, despite being a formally exact theory, in practical applications, FDE faces difficulties to describe situations in which a system is partitioned into strongly interacting subsystems, primarily due to the use of approximations to the kinetic energy density functional. 

Apart from standard response theory, FDE has also been used in combination with real-time TDDFT (rt-TDDFT) to obtain valence~\cite{de_santis_environmental_2020,de_santis_frozen-density_2022} and core excited states~\cite{de_santis_environment_2021}, though for discrete systems the response of the environment has, to date, been disregarded, preventing the description of the coupling of the response of all subsystems. In the context of condensed matter systems, subsystem couplings at rt-TDDFT have been taken into account, see~Ref.~[\citep{krishtal_subsystem_2015}].


In other QE approaches based on projection techniques, the difficulties in partitioning strongly interacting subsystems are effectively circumvented. One such technique, the Block-Orthogonalized Manby-Miller embedding (BOMME) method~\cite{ding_embedded_2017}, offers the advantage of dividing a system into two domains. The active region (Domain A) is treated with a high-level (mean-field) theory, while the remainder of the systems, taken to be the environment (Domain B), is treated with a lower level (mean-field) theory.  BOMME shows promise in reducing computational cost for standard electronic structure implementations that capture valence excited state energies for both intermolecular and intramolecular embedding schemes \cite{koh_development_2020}. Beyond ground-state properties, BOMME has been used to investigate valence\cite{koh_accelerating_2017} and core\cite{de_santis_environment_2021} excited states in combination with rt-TDDFT, and shown to be able to capture the coupling between the response of the different domains. However, these applications remain limited to molecular systems or clusters of small to moderate size, as with other projection-based approaches, the need to define basis functions spanning the whole system introduces a higher computational cost with respect to FDE.

In this work, we combine the strengths of BOMME (effective at capturing strong inter and intramolecular interactions) and FDE (efficient in including long-range interactions in large-scale systems) to accelerate real-time TDDFT calculations of core excited states, in what we shall refer to as multilevel rt-TDDFT quantum embedding. We will showcase this approach by carrying out simulations of the K-edge spectra of hydrated chloride and fluoride anions, as well as the {\edgelone}-edge spectra of hydrated chloride, complementing our prior work on valence~\cite{bouchafra_predictive_2018} and core~\cite{opoku_simulating_2022} ionizations.  In the case of chloride experimental K-edge spectra are available~\cite{antalek_solvation_2016}, together with TDDFT calculations on static structures where between 6 to 8 molecules surrounding the ion have been considered, each exhibiting a distinct spectral profile. Our first goal in this paper will be to revisit this question and assess how more sophisticated models compare to experimental results.



The manuscript is organized as follows: First, we provide a comprehensive review of the theoretical background, encompassing the formalism of rt-TDDFT, the theories of frozen density embedding and Block-Orthogonalized Manby-Miller embedding, along with their rt-TDDFT extensions, and their combination with the Multilevel embedding method presented here. Secondly, we discuss implementation details that enabled more efficient rt-TDDFT-FDE calculations. Thirdly, we present and thoroughly discuss our results obtained with different embedding methods, for the aqueous chloride and fluoride. Finally, we offer concluding remarks and outline potential avenues for further research.

\section{Theoretical Background \label{theory}}

Ground-state density functional theory (DFT) is governed by the Kohn-Sham (KS) equations expressed in atomic units as follows:
\begin{equation}
    \label{kseq}
    \left( -\frac{\nabla_{i}^2}{2} + v_{\mathrm{eff}}[\rho](\vec{r}) \right) \psi_{i}(\vec{r}) = \epsilon_{i} \psi_{i}(\vec{r}).
\end{equation}
In the above equation, we define the KS orbitals, $\psi_i(\vec{r})$, and the KS orbital energies, $\epsilon_{i}$. 

Time-dependent DFT, instead, is defined by the time-dependent KS equations which yield the time-dependent KS orbitals. Namely,
\begin{equation}
    \label{tdkseq}
    \left( -\frac{\nabla_{i}^2}{2} + v_{\mathrm{eff}}(\vec{r},t) \right) \psi_{i}(\vec{r},t) = i\frac{\partial \psi_{i}(\vec{r},t)}{\partial t}.
\end{equation}
The complexity arises in the time-dependence of the effective potential, which no longer solely depends on the electron density at time $t$ (i.e., $\rho(\vec{r},t)$). It also depends on the entire history of electron densities, the initial KS state, and the initial state of the interacting system \cite{vignale_real-time_2008,maitra_perspective_2016}.
 
Eq.\ (\ref{tdkseq}) can be rearranged to provide an equation of motion for the KS one-electron reduced density matrix (1-rdm). We will indicate by $\textbf{D}$ the KS 1-rdm expressed in some basis set as $\rho(\vec{r},t) = \sum_{\mu\nu}\textbf{D}_{\mu\nu}(t) \chi_\mu(\vec{r})\chi_\nu(\vec{r})$ (where $\chi$ are the chosen basis functions).

The time evolution of the KS density matrix can be formulated in terms of the \textit{Liouville-von Neumann}\cite{jakowski_liouvillevon_2009} (LvN) equation, expressed in an orthonormal basis\cite{li_time-dependent_2005} as
\begin{align}
	i\frac{\partial \textbf{D}(t)}{\partial t} = \textbf{F}(t) \textbf{D}(t) -  \textbf{D}(t) \textbf{F}(t),
\end{align}
which depends on the \textit{time-dependent Fock matrix}, $\textbf{F}(t)$. During the propagation, the matrix elements $\textbf{F}(t)$ are calculated in the atomic orbital (AO) basis. In our implementation, driven by Psi4\cite{turney_psi4_2012} and Psi4Numpy\cite{smith_p_2018} frameworks, the AO basis set consists of atom-centered Gaussian-type functions, or GTOs.

The solution of the LvN equation is not straightforward for two reasons. First, in principle the dependency of $\textbf{F}(t)$ on $\textbf{D}$ is very complicated and non-adiabatic. We avoid such complications by applying the adiabatic approximation and assuming that the dependency is simply given by $\textbf{F}(t) = \textbf{F}[\textbf{D}(t)]$. Second, even in the adiabatic approximation, the dependency of $\textbf{F}$ on $\textbf{D}$ is highly nontrivial and includes a non-linearity that needs to be taken into account during time evolution. Thus, to handle this non-linearity, we employ a propagator approach:
\begin{equation}
	\textbf{D}(t) = \textbf{U} (t,t_0) \textbf{D}(t) \textbf{U} (t,t_0)^\dagger.
\end{equation}
Here, $\textbf{U} (t,t_0)$ is the time evolution operator.  To define the propagator, we first need to define the Fock matrix. For a non-relativistic Hamiltonian the Fock matrix, $\textbf{F}(t)$, is expressed as
	\begin{equation}\label{fock_time}
		\textbf{F}(t) = \textbf{T} + \textbf{V}_{eN} + \textbf{V}_{xc}[\textbf{D}(t)] + \textbf{J}[\textbf{D}(t)]+ \textbf{V}_{ext}(t).
	\end{equation}
Here, $\textbf{T}_{\mu\nu}=\langle\chi_\mu\left|\frac{1}{2}\nabla^2\right| \chi_\nu\rangle$ is the single-particle kinetic energy matrix, $\textbf{V}_{eN}$ is the matrix associated with the local electron-nuclear attraction potential, $v_{eN}(\vec{r})$. The exchange-correlation potential, $v_{xc}(\vec{r})$ and the classical electron-electron Coulomb repulsion, are represented by the matrices $\textbf{V}_{xc}$ and $\textbf{J}$, respectively. 

In the adiabatic approximation, the explicit time dependence in the Fock matrix arises not only from the time-dependent density, $\textbf{D}(t)$, in the Coulomb, $\textbf{J}[\textbf{D}(t)]$, and exchange-correlation, $\textbf{V}_{xc}[\textbf{D}(t)]$, potentials but also from the contribution of the time-dependent external potential, given in matrix form as $\textbf{V}_{ext}(t)$ (e.g., an applied laser field). 
	
The time evolution operator is given by
	\begin{align*}
		\textbf{U}(t,t_0) = \hat{\mathcal{T}} \exp\left( -i \int_{t_0}^{t} \textbf{F} (t') dt'  \right).
	\end{align*}

The time-ordering operator $\hat{\mathcal{T}}$ arises due to the fact that $\textbf{F} (t)$ at different times do not necessarily commute ($[\textbf{F} (t), \textbf{F} (t')] \ne 0$). Sensible approximations of arbitrary precision can be made by discretizing the propagation time into short time steps. Among various propagation schemes \cite{press_numerical_2007,meng_real-time_2008}, here, we adopt the second-order midpoint Magnus propagator\cite{magnus_exponential_1954,casas_explicit_2006,li_time-dependent_2005}:
	\begin{align*}
		\textbf{U}(t+\Delta t, t) \approx \exp \left[-i \textbf{F}\left(t+\frac{\Delta t}{2}\right) \Delta t\right].
	\end{align*} 

\subsection{Frozen density embedding in rt-TDDFT}
	
In frozen density embedding (FDE), the electron density of the total system is partitioned into subsystem electron densities, $\{\rho_I(\vec{r})\}$, and is written as, $\rho_{tot}(\vec{r}) = \sum\limits_{I=1}^{N_S} \rho_{I}(\vec{r})$, where $N_S$ accounts for the total number of subsystems. 
	
For simplicity, we consider a partition of the total density into only two contributions:
	\begin{equation}\label{eq:sumdensity}
		\rho_{tot}(\vec{r}) = \rho_{1}(\vec{r}) + \rho_{2}(\vec{r}).
	\end{equation}
The total electronic energy is defined as,
	\begin{align*}
		E_{tot}[\rho_1,\rho_{2}]= E_1[\rho_{1}] + E_{2}[\rho_{2}] + E_{int}[\rho_{1},\rho_{2}],
	\end{align*}
where we introduce an interaction energy $E_{int}$ which is formally bifunctional of the two subsystem densities. The subsystem energies are defined as:
    \begin{equation}
		E_{I}[\rho_{I}] = T_{s}[\rho_{I}] + E_{NN}^I + \int v_\mathrm{eN}^I(\vec{r})\rho_I(\vec{r})d\vec{r} + E_H[\rho_{I}] + E_{\mathrm{xc}}[\rho_{I}],
	\end{equation}
where $v_\mathrm{eN}^I(\vec{r})$ denotes the electron-nuclear attraction potential of subsystem $I$. While $E_H[\rho_I]$ and $E_{\mathrm{xc}}[\rho_{I}]$ are the Hartree and exchange-correlation energy functionals, and $E_{NN}^I$ is the nuclear repulsion energy of subsystem $I$. And $T_{s}[\rho_{I}]$ is the subsystem non-interacting kinetic energy. 
	
The interaction energy is defined as,
	\begin{align}
		\label{fde}
  \nonumber
		E_{int}[\rho_{1},\rho_{2}] = T_{s}^{nadd}[\rho_{1},\rho_{2}] + E_{NN}^{1,2} + \int \rho_{1}(\vec{r}) v_\mathrm{eN}^{1}(\vec{r})d\vec{r} + \int \rho_{2}(\vec{r}) v_\mathrm{eN}^{2}(\vec{r})d\vec{r} + \\
		\iint\frac{\rho_{1}(\vec{r})\rho_{2}(\vec{r'})}{|\vec{r}-\vec{r'}|} d\vec{r}d\vec{r}^\prime+ E_{\mathrm{xc}}^{nadd}[\rho_{1},\rho_{2}] ,
	\end{align}
	where the non-additive kinetic energy, $T_{s}^{nadd}[\rho_{1},\rho_{2}]=T_{s}[\rho_{tot}]-T_{s}[\rho_{1}]-T_{s}[\rho_{2}]$, and non-additive exchange-correlation, $E_{\mathrm{xc}}^{nadd}[\rho_{1},\rho_{2}]=E_{\mathrm{xc}}[\rho_{tot}]-E_{\mathrm{xc}}[\rho_{1}]-E_{\mathrm{xc}}[\rho_{2}]$, arise because they are not linear functionals of the density. Finally, $E_{NN}^{1,2}$ is the nuclear-nuclear repulsion of nuclei in subsystem 1 with the ones in subsystem 2.
	
The variational problem is given by the set of two KS equations with constrained electron density \cite{wesolowski1994ab}, which read for subsystem 1:
\begin{align}\label{ksembpot}
	\left( -\frac{\nabla_{i}^2}{2} + v_{\mathrm{eff}}^{1}[\rho_1](\vec{r}) + v_{\mathrm{emb}}^{1}[\rho_{1},\rho_{2}](\vec{r}) \right) \psi_{i}^1(\vec{r}) = \epsilon_{i}^1 \psi_{i}^1(\vec{r}),
\end{align}
where $v_{\mathrm{eff}}^{1}[\rho_1](\vec{r})$ is the subsystem 1 effective KS potential. The interaction of the electrons of subsystem 1 with the environment in FDE is represented by an embedding potential, 
\begin{align}\label{embpot}
		v_{\mathrm{emb}}^{1}[\rho_1,\rho_{2}](\vec{r}) = v_{eN}^{2}(\vec{r})  + \int\frac{\rho_{2}(\vec{r'})}{|\vec{r}-\vec{r'}|}d\vec{r'} + v_{\mathrm{xc}}^{nadd}[\rho_1,\rho_{2}](\vec{r})  + v_{\mathrm{T_s}}^{nadd}[\rho_1,\rho_{2}](\vec{r}) .
\end{align}
Where,
    
\begin{align}
\label{non_def1}
        v_{\mathrm{xc}}^{nadd}[\rho_1,\rho_{2}](\vec{r})  =&  \frac{\delta E_{xc}[\rho_{tot}]}{\delta \rho_{tot}(\vec{r})} -  \frac{\delta E_{xc}[\rho_{1}]}{\delta \rho_{1}(\vec{r})},\\
\label{non_def2}
        v_{\mathrm{T_s}}^{nadd}[\rho_1,\rho_{2}](\vec{r})  =&  \frac{\delta T_s[\rho_{tot}]}{\delta \rho_{tot}(\vec{r})} -  \frac{\delta T_s[\rho_{1}]}{\delta \rho_{1}(\vec{r})}.
	\end{align}
	
Due to the direct dependence of the non-additive terms on both subsystem densities, $\rho_{tot}$, an approximation for the exchange-correlation and kinetic energy functionals and potentials must be employed. Pure density functionals are used, providing a clear and simple way to evaluate all non-additive terms. Similar to the exchange-correlation functional approximants, there are several options for the kinetic energy, such as the Thomas-Fermi \cite{thomas_calculation_1927} kinetic energy functional, or other GGA-based functionals \cite{laricchia2011generalized,lembarki_obtaining_1994}. 
	
The set of coupled equations, Eq.\ (\ref{ksembpot})  and Eq. \ (\ref{embpot}), can be solved iteratively using a Freeze-and-Thaw\cite{FaT} procedure, which is repeated until all the subsystems' densities converge. Once the embedding potential is obtained, $v_{\mathrm{emb}}^{1}[\rho_1,\rho_{2}](\vec{r})$, it is added to the usual Fock Hamiltonian, by computing its matrix elements in the atomic orbitals basis. The $\mu,\nu$ element is
\begin{equation}\label{matrix_vemb}
		\left(\textbf{V}_{\mathrm{emb}}^1\right)_{\mu \nu} = \int \chi_{\mu}(\vec{r}) v_{\mathrm{emb}}^{1}[\rho_1,\rho_{2}](\vec{r})\chi_{\nu}(\vec{r}) d\vec{r}.
\end{equation}
	
The uncoupled real-time FDE TDDFT scheme (\textit{uFDE-rt-TDDFT}), in which  $\rho_{2}$ is kept frozen and therefore does not respond to the external perturbation, is implemented \cite{de_santis_environmental_2020} by adding the matrix representation of the embedding potential, Eq.\ (\ref{matrix_vemb}), to the time-dependent Fock matrix, Eq.\ (\ref{fock_time}), for the ``active'' subsystem 1, which reads as, 
\begin{align}\label{fock_emb}
	\textbf{F}^{1}(t) = \textbf{T}^{1} + \textbf{V}^{1}_{nuc} + \textbf{V}_{xc}[\rho_{1}(t)] + \textbf{J}[\rho_{1}(t)]+ \textbf{V}_{ext}(t) + \textbf{V}_{emb}[\rho_{1}(t),\rho_{2}],
\end{align}	
where we have included explicitly the functional dependence of each term.
    
Clearly, a time dependence is added to the embedding potential, which needs to be updated during the propagation, even though the density of the environment, $\rho_{2}$, is taken to be frozen.
	
\subsection{Block-Orthogonalized Manby-Miller embedding in rt-TDDFT}
	
Block-Orthogonalized Manby-Miller embedding (BOMME) \cite{manby_simple_2012,pipolo_cavity_2014} is a mean-field embedding approach that partitions the total system into two different domains, A and B, treated with two different levels of theory. In this method, a high-level Fock matrix is assigned to the domain requiring accurate treatment (usually A), while the remaining part of the system (domain B) is described by a low-level Fock matrix. To accomplish this, instead of using the conventional atomic-orbital (AO) basis, the high and low levels of theory are encoded by a block-orthogonal (BO) basis. Thus, a basis set transformation matrix \textbf{O} is formulated,
\begin{equation}
\textbf{O}=
	\begin{pmatrix}
		\textbf{I}^{AA} & -\textbf{P}^{AB} \\
		\textbf{0} & \textbf{I}^{BB}
	\end{pmatrix}.
\end{equation}
 
This matrix explores the projection of the non-orthogonal AO basis set into the BO basis set. Here the diagonal terms, $\textbf{I}^{AA}$ and $\textbf{I}^{BB}$, account for the identity matrices of domains A and B, respectively, with dimensions corresponding to the sizes of their respective basis sets ($n_a$ and $n_b$). While, $\textbf{P}^{AB} = (\textbf{S}^{AA})^{-1}\textbf{S}^{AB}$ corresponds to the projection matrix, where $\textbf{S}^{AB}$ is the AO overlap between the atomic orbitals assigned to domains A and B. 
  
By considering AA(BB) block to denote the domain approached with a high-(low-) level of theory, the \textit{time-dependent} Fock matrix in the BO basis reads as,
\begin{equation}\label{fock_bomme}
	\textbf{F}(t)=\mathbf{\tilde{h}}_0 + \mathbf{\tilde{G}}^{Low}[\mathbf{\tilde{D}}(t)] + ( \mathbf{\tilde{G}}^{High}[\mathbf{\tilde{D}}^{AA}(t)] -  \mathbf{\tilde{G}}^{Low}[\mathbf{\tilde{D}}^{AA}(t)] ) + \textbf{V}_{ext}(t),
\end{equation}
where $\mathbf{\tilde{h}}_0$ correspond to the one-electron operator, while $\textbf{G}[\mathbf{\tilde{D}}(t)]$ denoted the two-electron terms in the BO basis. The foregoing is obtained by the following transformations,	\begin{align}
\mathbf{\tilde{h}}_0 = \textbf{O}^T \ \mathbf{h}_0 \ \textbf{O} \\
	\tilde{\textbf{G}}[\mathbf{\tilde{D}}(t)]^{High/Low} = \textbf{O}^T\ \textbf{G}[\textbf{D}(t)]^{High/Low} \ \textbf{O} \\
	\tilde{\textbf{D}}(t) =  \textbf{O}^T \ \textbf{D}(t)  \ \textbf{O},
\end{align}
where $\textbf{h}_0$ and $\textbf{G}[\textbf{D}(t)]$ corresponds to the one and two electrons terms in the AO basis,
\begin{align}
	\textbf{h}_0 = & \textbf{T} + \textbf{V}_{eN}\\
	\textbf{G}[\textbf{D}(t)] = & \textbf{J}[\textbf{D}(t)] + c_x \textbf{K}[\textbf{D}(t)] + c_x\textbf{V}_{xc}[\textbf{D}(t)],
\end{align}
and $c_x$ is the fraction of the exact Hartree-Fock exchange in the exchange-correlation potential $\textbf{V}_{xc}$. The exchange term in $\textbf{G}[\textbf{D}(t)]^{High}$ is calculated by considering only the exact exchange interaction within the AA block \cite{koh_accelerating_2017}. This method is called \textit{BOMME-rt-TDDFT}.
	
\subsection{Multilevel embedding in rt-TDDFT}

Up to now, we have described two types of embedding methods. One, FDE, splits a system into two or more subsystems which are then allowed to interact via orbital-free non-additive functionals. The second method, BOMME, splits a system into domains at the density matrix level, using a basis set representation. BOMME then treats one domain (called A) with one DFT method (called $High$), and the other domain (called B) with another DFT method (called $Low$). In this work, we combine the two approaches in a single method called multilevel embedding rt-TDDFT or ML-rt-TDDFT. This integration ensures that strongly interacting fragments in a system are treated with BOMME, while weakly interacting fragments are managed with FDE. By doing so, we leverage the excellent behavior of BOMME for reproducing density and energy of strongly-interacting fragments \cite{de_santis_environmental_2020,ding_embedded_2017} and that of FDE for reproducing density and energy of weakly-interacting fragments~\cite{Gomes2008,Hfener2012}. ML-rt-TDDFT completely avoids a KS-DFT calculation of the full system affording massive computational savings. In practice, ML-rt-TDDFT prescribes the solution of Kohn-Sham-like equations in which an FDE embedding potential is added to the BOMME Fock matrix, both with and without the external perturbation. With that, the time-dependent Fock matrix in Eq.\ (\ref{fock_bomme}) reads
\begin{align*}
	\textbf{F}(t)=\mathbf{\tilde{h}}_0 + \mathbf{\tilde{G}}^{Low}[\mathbf{\tilde{D}}(t)] + ( \mathbf{\tilde{G}}^{High}[\mathbf{\tilde{D}}^{AA}(t)] -  \mathbf{\tilde{G}}^{Low}[\mathbf{\tilde{D}}^{AA}(t)] ) + \textbf{V}_{ext}(t) + \textbf{V}_{emb}^{1}[\rho_{1}(t),\rho_{2}].
\end{align*}

With such a definition, one introduces an intermediate layer (the low-level BOMME region, or domain B) between domain A (e.g.\ a solute, in our case the halide ions) and the FDE environment. This reduces computational cost in treating the entirety of subsystem 1 with the high-level of theory. While, in principle, one could employ FDE to treat domains A and B as two additional subsystems, the use of BOMME instead allows for a more reliable treatment of stronger interactions within the domains, since the use of approximate non-additive kinetic energy density functionals (NAKEs), which are unable to capture strong interactions \cite{konecny_acceleration_2016,repisky_excitation_2015} is avoided. It also alleviates the storage burden associated with employing a common basis set for the whole system. Considering the long propagation times necessary to simulate core spectra, reducing computational complexity and storage requirements is particularly welcome. 

It should be noted that in the prior \textit{uFDE-rt-TDDFT} implementation~\cite{de_santis_environmental_2020} (see text after Eq.\ (\ref{matrix_vemb})),  due to implementation issues, the computational time required to evaluate the embedding potential $v_{emb}[\rho^{1}(t),\rho_{2}](\vec{r})$ in  Eq.\ (\ref{embpot}), in the uncoupled scheme at each time iteration was found to increase linearly with the number of water molecules surrounding the solute. This increase poses a challenge, with the computation of Eq.\ (\ref{matrix_vemb}) being the most time-consuming step. This required the use of a step-wise updating strategy for calculating the embedding potential matrix $\textbf{V}_{emb}[\rho_{1}(t),\rho_{2}]$, which showed promising performance when updated every 30~steps throughout the simulation time. However, this approach can become problematic due to variations in the density of the active system, seriously compromising its use in further production calculations.


	
To address the efficiency issue in evaluating the embedding potential, Eq.\ (\ref{embpot}), we have explored the reimplementation of two crucial PyBertha functions (namely \textit{embpot2mat.py} and \textit{denstogrid.py} from the \textit{psi4embrt.py} module) (Referred as ``partial'' GPU-aware implementation) employing two optimize tensor libraries, PyTorch\cite{imambi_pytorch_2021} and TensorFlow\cite{tensorflow_developers_tensorflow_2021}, in view of their ability to offload matrix operations to GPUs while remaining very similar to the Numpy API (thus requiring rather localized changes to the Numpy-based prior implementation). We have also reimplemented all Numpy matrix operations (Referred as ``full'' GPU-aware implementation) of the \textit{util.py} function of the psi4rt module along with the aforementioned PyBertha functions. This was done because the \textit{mo\_fock\_mid\_forwd\_eval.py} of \textit{util.py} continuously evaluates the Hatree-Fock, Exchange, and Coulomb matrices by adding the analytical $\delta$-function pulse at each iteration along the real-time propagation. Further implementation details are available in the supplementary material (Section~\ref{sup-GPU-uFDE-rt-TDDFT}).

\begin{figure}[!htbp]
    \captionsetup[subfigure]{labelformat=empty}
    \centering
    \begin{subfigure}{\linewidth}
        \includegraphics[width=\linewidth]{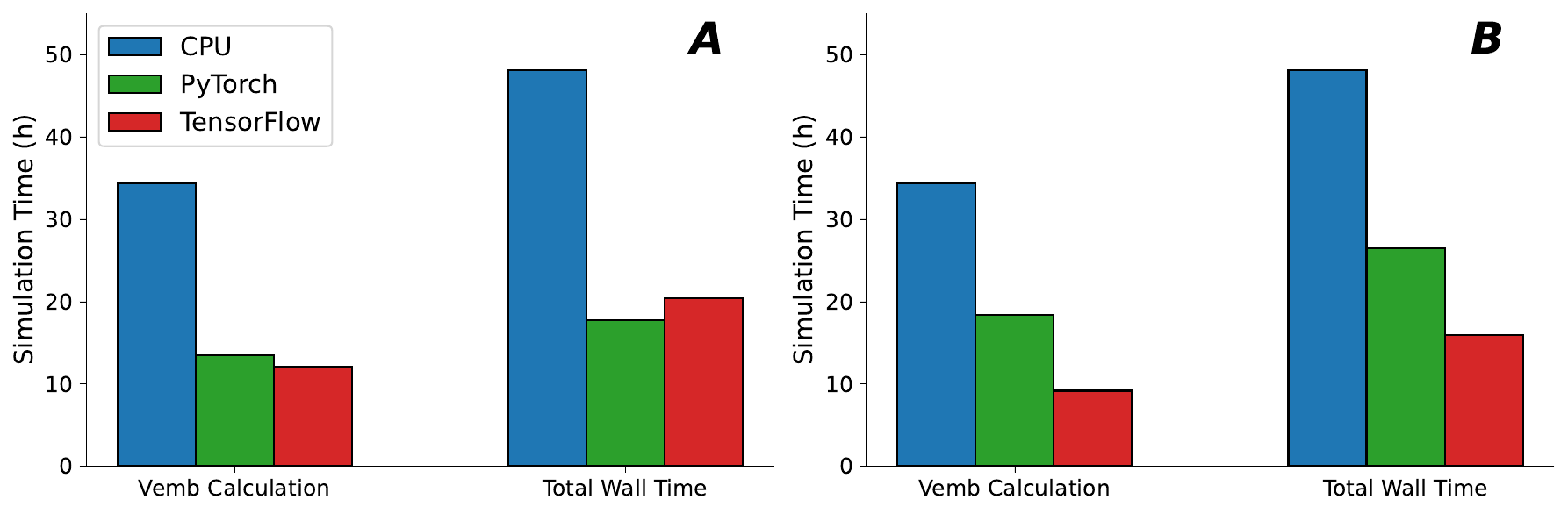}
    \end{subfigure}
    \caption{Time profile of GPU implementation. Insight A:	Full implementation in \textit{psi4rt.py} module along with the aforementioned PyBertha functions (Full GPU implementation). Insight B: Total time in seconds spent by the main Python functions (\textit{embpot2mat.py} and \textit{denstogrid.py}) involved in the evaluation of Eq.~\ref{embpot} in \textit{psi4embrt.py} module of PyBertha code (Partial GPU implementation). The time corresponds to the total time spent in the function per call. The total number of calls of both functions is \num{29007}. The percentages are obtained by cProfile and Snakeviz\cite{davis2021snakeviz} packages. The total simulation time is \qty{33.9}{\femto\second}, for a total of \num{56000} steps, with a length of \num{0.025}~a.u. per time step.}
    \label{timings}
    \end{figure}

The results of the total timing for a single snapshot (SN-14 of chloride embedded in its first solvation shell) using both ``full'' and ``partial'' implementations are displayed in figures A and B of Fig.~\ref{timings}, respectively. By using PyTorch and TensorFlow, we reduced simulation time from 48 to 16 and 26 hours, respectively when the partial GPU implementation is used. Interestingly, the embedding potential calculations (referred to as the \textit{``Vemb calculation''}) are three times faster with TensorFlow than with the conventional CPU implementation, and almost twice as fast with PyTorch when partial implementation is used, Figure B in  Fig. \ \ref{timings}. Surprisingly, when we include the full optimization, Figure A in  Fig. \ \ref{timings}, including \textit{psi4rt.py} functions, the TensorFlow time increases while the PyTorch time decreases, thus in the ''Partial'' TensorFlow the embedding potential evaluation (\textit{``Vemb calculation''}) differs by 10 hours with respect to partial PyTorch implementation.

\begin{figure}[!htbp]
    \captionsetup[subfigure]{labelformat=empty}
    \centering
    \begin{subfigure}{1.0\linewidth}
        \includegraphics[width=\linewidth]{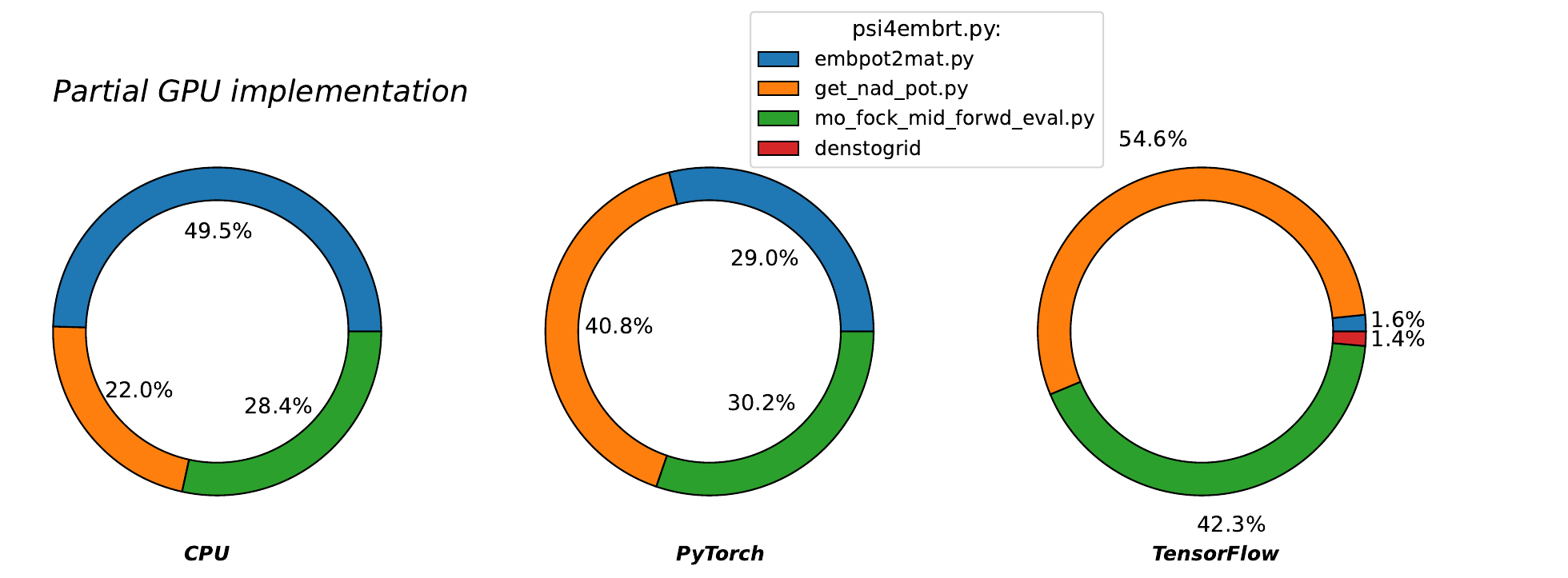}
    \end{subfigure}
    \begin{subfigure}{1.0\linewidth}
        \includegraphics[width=\linewidth]{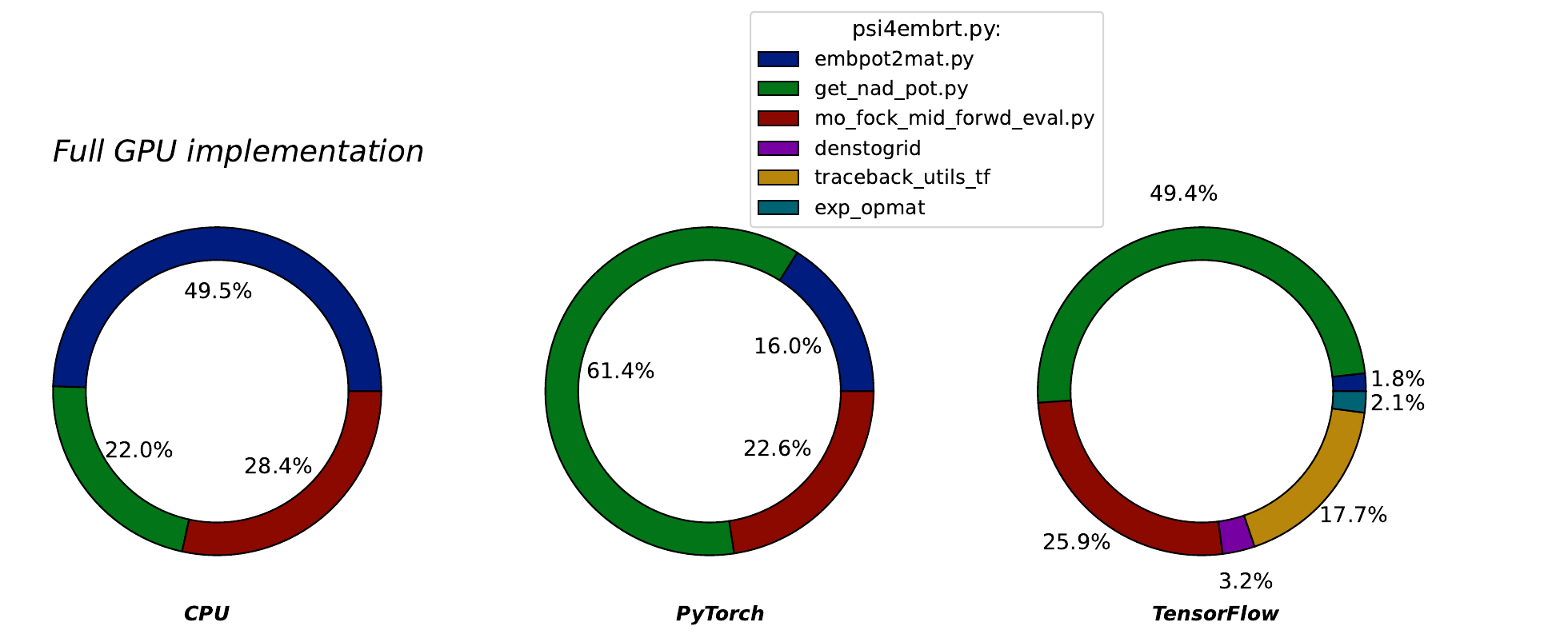}
    \end{subfigure}
    \caption{Profiling results for the main Python functions (\textit{embpot2mat.py} and \textit{denstogrid.py}) involved in the evaluation of Eq. \ref{embpot} in \textit{psi4embrt.py} module of PyBertha code (Partial GPU implementation). And the full implementation in \textit{psi4rt.py} module along with the aforementioned PyBertha functions (Full GPU implementation). The percentages are obtained by cProfile and Snakeviz\cite{davis2021snakeviz} packages. An interactive version of the Profile can be found in the Zenodo repository~{\zenodo}.}
    \label{profile}
\end{figure}

The above can be rationalized by looking at basic profiling results for the module \textit{psi4emb\-rt.py} for both ``partial'' and ``full'' GPU implementations, Fig. \ref{profile}. These profiling results indicate that \qty{49.5}{\percent} of the total time is spent in the \textit{embpot2mat.py} module when a CPU-only (Numpy-based) implementation is used, left-hand side figure in Fig. \ref{profile}. In the TensorFlow and PyTorch partial implementations, the time spent on the same routine, \textit{embpot2mat.py}, decreases to \qty{1.6}{\percent} and \qty{29.0}{\percent}, respectively. This is not the case when the ``full'' implementation is profiled where the time spent in the \textit{embpot2mat.py} is reduced to \qty{16.0}{\percent} in the PyTorch case and \qty{1.8}{\percent} in the TensorFlow case. We attribute these differences to the continuous transfer of data between Numpy arrays on the CPU and TensorFlow tensors on the GPU, which is known to negatively affect the performance of TensorFlow\cite{michelucci_tensorflow_2019}. This is especially true when transitioning from partial to full implementation, as the latter involves more matrix-tensor transformations. Despite this, we insist on the main point of employing optimized tensor libraries, which is to provide significant performance improvements with relatively minor changes to the Numpy-based code. In subsequent optimizations of the code work, we plan to better characterize this point, while attempting to eliminate other performance bottlenecks.

\clearpage
\section*{Computational details\label{comp}}

\textit{ML-rt-TDDF} real-time framework, was implemented as part of PyBertha \cite{foster_bertha_2020,de_santis_pyberthart_2020,belpassi_bertha_2020} package in the \textit{numercialtest} branch of PyBertha version 64f6752\cite{Pybertha_nt}. PyBertha is a code based on Psi4Numpy\cite{smith_p_2018} framework, here we used  Psi4\cite{smith_p_2020} version 1.8a1.dev5. \textit{BOMME-rt-TDDFT} real-time code, also coupled to the PyBertha package is available in the GitHub repository (version ba72aec\cite{BOMME-rt}). Examples of user usage are published in the Zenodo repository~{\zenodo}.

We modified the previously published version\cite{de_santis_environmental_2020} of \textit{uFDE-rt-TDDFT} in PyBertha repository, by re-implementing the projection of the $\left(\textbf{V}_{\mathrm{emb}}^i\right)_{\mu \nu}$ into the AO basis, see Eq. \ref{matrix_vemb} with PyTorch 1.13.1\cite{imambi_pytorch_2021} (code available in the Git-Hub revision c3e1a57 \cite{Psi4rt-FDE-pytorch}) and TensorFlow 2.13.0\cite{tensorflow_developers_tensorflow_2021} (code available in the Git-Hub revision 88092d1 \cite{Psi4rt-FDE-tensorflow}).
	
In \textit{uFDE-rt-TDDFT} approach, we exmployed two schemes, where halides served as active subsystems (subsystem 1), and were embedded in 1) \num{8}~water molecules; and 2) \num{50}~water molecules, as subsystem 2. For subsystem 1, we employed the B3LYP functional\cite{becke_density-functional_1993} with an aug-cc-pVTZ\cite{papajak_perspectives_2011} basis set. The embedding potential of the active subsystem for the remaining water molecules was calculated employing BLYP functional and a DZP basis set\cite{dunning_gaussian_1970}. The benchmarking was carried out using 4 OpenMP threads and 1 GPU NVIDIA A100-PCIE-40GB device.
We also investigated the effect of the relaxation of the density of the environment, by implementing the Freeze and thaw (FaT) procedure of PyADF \cite{jacob_pyadf_2011} into \textit{uFDE-rt-TDDFT} and  \textit{ML-rt-TDDF} codes. Thus, one can start the time-propagation with environment ground-state densities which are either ``relaxed'' (from a prior FaT calculation with \num{5}~cycles) or ``unrelaxed''. 
	
In \textit{ML-rt-TDDFT}, the halides (domain A) and their first solvation shell (8 water molecules) (domain B) made up the active subsystem or subsystem 1, while the remaining water molecules (up to 50) belonged to the environment subsystem, subsystem 2. For the active subsystem, the halide was calculated at B3LYP, while for the surrounding waters up to 8 molecules, we used BLYP as exchange-correlation functional, with the aug-cc-pVDZ\cite{papajak_perspectives_2011} and DZP basis sets, respectively. To include the embedding potential $\left(\textbf{V}_{\mathrm{emb}}^i\right)_{\mu \nu}$ of the remaining \num{42}~waters in \textit{ML-rt-TDDF}, we used ADF\cite{te_velde_chemistry_2001} version 2019.403, employing the PyADF \cite{jacob_pyadf_2011} code. Environment subsystem calculations were carried out with a DZP \cite{van_lenthe_optimized_2003} Slater Type Orbitals basis set that differs from the one used by the Psi4 package (Gaussian Type Orbitals) at the BLYP level. The setup for the active subsystem or subsystem 1, in \textit{ML-rt-TDDFT} described above is employed in \textit{BOMME-rt-TDDFT} calculations, with the halide being Domain A (B3LYP/aug-cc-pVDZ), and first solvation shell Domain B (BLYP/DZP) but with a Gaussian Type Orbital base package.
	
In all the real-time simulations, the initial electron density $\delta_0$ is perturbed by an analytical $\delta$-function pulse with a strength of $\kappa=5.0x10^{-5}$ a.u. along x, y, and z directions. The induced dipole moment was kept in each iteration for a total of \num{56000}~steps, with a length of \num{0.025}~a.u. per time step. Thus, the total simulation time is \qty{33.9}{\femto\second} for both halides. Similarly, as previous works\cite{de_santis_environmental_2020,de_santis_environment_2021}, we employed a Padé approximant-based Fourier Transform (FFT) \cite{bruner_accelerated_2016} with an exponential damping $e^{-\lambda t}$ with $\lambda=$\num{3.0e-4}~a.u..
	
\begin{figure}[!htbp]
    \captionsetup[subfigure]{labelformat=empty}
    \centering
    \begin{subfigure}{0.45\linewidth}
        \includegraphics[width=\linewidth]{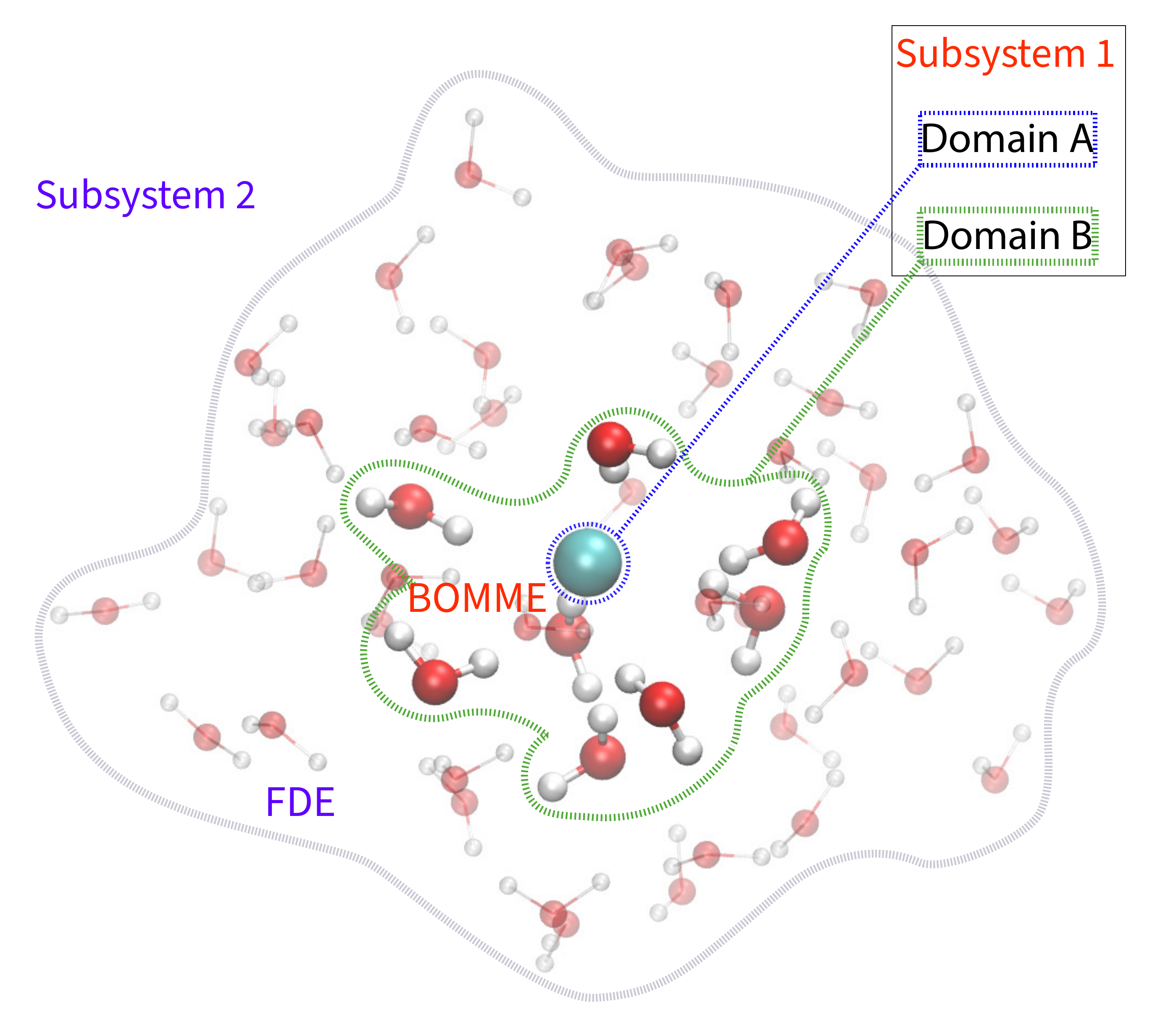}
    \end{subfigure}
    \begin{subfigure}{0.45\linewidth}
        \includegraphics[width=\linewidth]{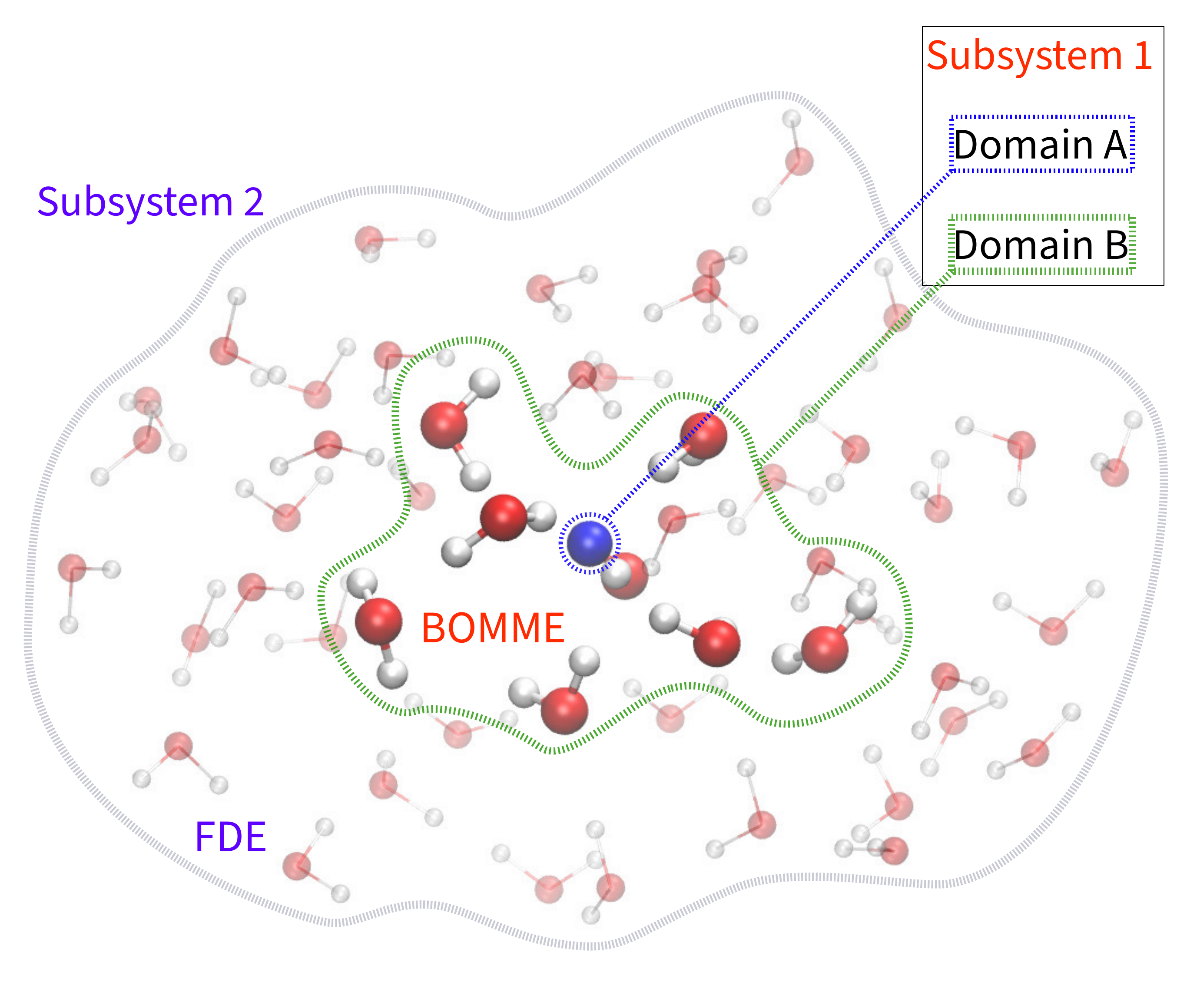}
    \end{subfigure}
    \caption{Schematic representation of the multilevel approach for embedding chloride (Right) and fluoride (Left) anions in up to \num{50}~water molecules. Subsystem 1 corresponds to the set of Domain A: Halides (\ce{Cl-}, and \ce{F-}) and Domain B: water first solvation shell (up to \num{8}~water molecules). Subsystem 2 corresponds to the second or outer shell (up to \num{50}~water molecules)}
    \label{systems}
    \end{figure}

All real-time simulations were carried out utilizing \num{10}~configurations for both the fluoride and chloride systems. These were selected from snapshots of \qty{10}{\nano\second}-long classical molecular dynamics trajectories of halides in water carried out by~\citet{real_structural_2016} and cut out to build \num{50}~water droplets as used in the study by~\citet{bouchafra_predictive_2018} (see Fig.~\ref{systems}). Out of these structure, we selected five snapshots based on the water binding energies (BE) predicted by employing the relativistic equation-of-motion coupled-cluster and FDE\cite{bouchafra_predictive_2018}. Specifically, we considered the geometries with the closest BEs concerning the average binding energy of the $1b_1$ peak. Similarly, we choose the other five geometries based on the closest representative geometries concerning the binding energy distribution of the corresponding halide.

We utilized an ensemble approach \cite{zuehlsdorff_modeling_2019} which involved averaging over multiple core excitations obtained from \num{10}~snapshots per system. In the ensemble approach, the excitation peaks obtained after the Fourier transform (FFT) were convoluted with a Gaussian function of width $\sigma$=\qty{0.7}{\electronvolt}, which was centered at each excitation. The final step involved summing the convoluted peaks to create the absorption spectrum.

\section{Results and discussion\label{results}}
    
\subsection{\ce{[X(H2O)8]-} models}

Before discussing the performance of \textit{ML-rt-TDDFT} calculations on the \ce{[X(H2O)50]-} models and its comparison to experiment, it is instructive to explore the behavior of uncoupled FDE and BOMME for the smaller \ce{[X(H2O)8]-} models (Subsystem 1). Some of us have previously carried out~\cite{de_santis_environment_2021} a similar comparison for the K-edges of chloride and fluoride, and the {\edgelone}-edge of chloride. In this context, the same general conclusions persist regarding the relative merits and shortcomings of FDE and BOMME. However, However, the prior calculations only explored a single structure, whereas here, we extend the analysis to consider the effect of configurational averaging on the calculated spectra. This discussion is crucial as it will later contribute to the exploration of the impact of outer solvation shells on the spectra. Additionally, we systematically investigate the effect of the relaxation of the ground-state density of the environment (after several FaT cycles) in FDE calculations in a more systematic manner than in previous work.

\subsubsection{uFDE rt-TDDFT}\label{sec-ufde}

\begin{figure}[!htbp]
    \captionsetup[subfigure]{labelformat=empty}
    \centering
    \begin{subfigure}{0.75\linewidth}
        \includegraphics[width=\linewidth]{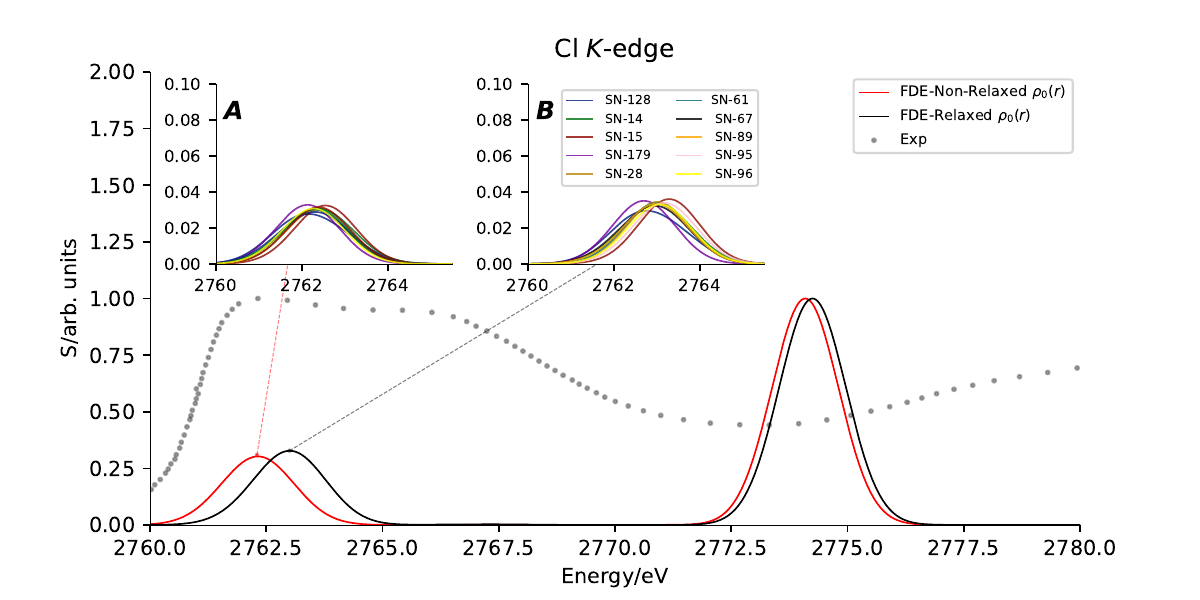}
    \end{subfigure}\hspace{0.1em}
    \begin{subfigure}{0.75\linewidth}
        \includegraphics[width=\linewidth]{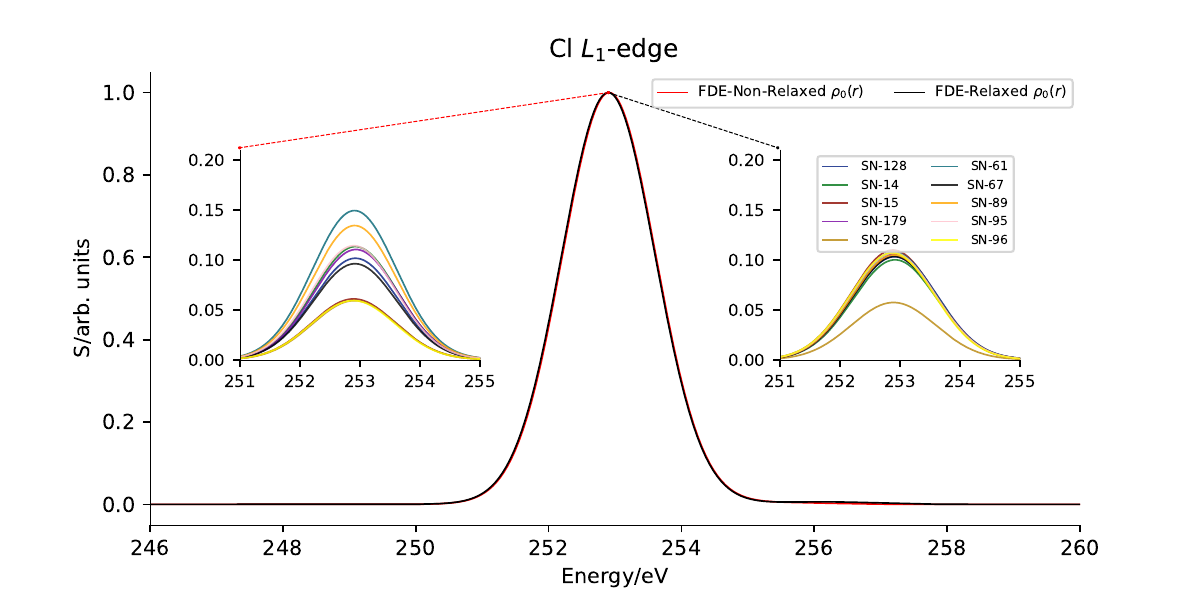}
    \end{subfigure}\hspace{0.1em}
    \begin{subfigure}{0.75\linewidth}
        \includegraphics[width=\linewidth]{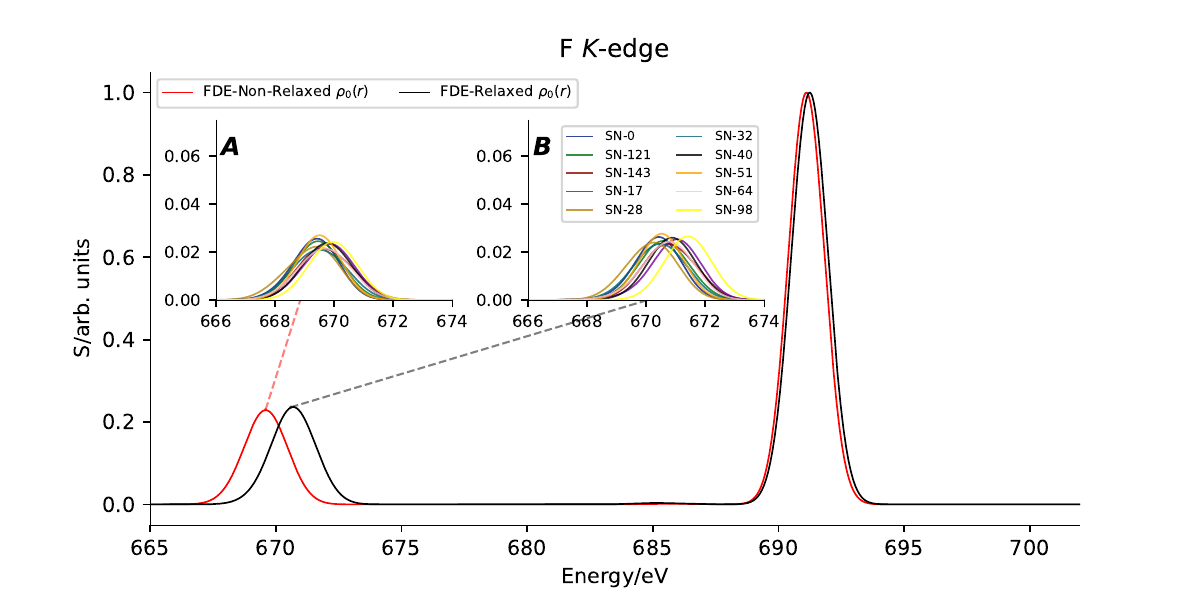}
    \end{subfigure}
    \caption{Chloride {\edgek}-edge (top) and {\edgelone}-edge (middle) as well as fluoride~{\edgek}-edge (bottom) X-ray absorption spectra for the anions embedded in their first solvation shell (\num{8}~water molecules) between the energy range of the free ion edge peak and up to \qty{20}{\electronvolt} (Cl {\edgek}-edge), \qty{14}{\electronvolt} (Cl {\edgelone}-edge), and \qty{35}{\electronvolt} (F {\edgek}-edge). Data was obtained by employing \textit{FDE-rt-TDDFT} with a non-relax and relaxed initial density $\rho (r)$. 5 cycles of Freeze and Thaw were employed to obtain a relaxed initial density. Insect pictures show the contribution per snapshot. A Gaussian broadening ($\sigma$ = \qty{0.7} {\electronvolt} was used.}
    \label{Spectra_Cl_LCL_F_FDE}
    \end{figure}

The \textit{uFDE-rt-TDDFT} simulated XAS spectrum for chloride, using ``unrelaxed'' and ``relaxed'' ground-state densities $\rho_0$(r) is shown in Fig.~ \ref{Spectra_Cl_LCL_F_FDE}. Starting with the K-edge, the first feature peak over \qty{2762}{\electronvolt} relates to the electronic transition from the 1s~core state to (n+1)p~valence state of the chloride anion. Similarly, the second peak can be assigned to the 1s~$\rightarrow$~(n+2)p transition~\cite{de_santis_environment_2021}. We observe a solvatochromic blue shift of~\qty{0.69}{\electronvolt}for the first and~\qty{0.16}{\electronvolt} for the second absorption peak, respectively, for the ``relaxed'' calculations with respect to the ``unrelaxed'' ones. This underscores the higher sensitivity of the orbitals in the valence region (including the lower-lying (n+1)p~virtuals) to the relaxation of the density~\cite{bouchafra_predictive_2018}; this is expected since we are dealing with a charged species that will strongly polarize the first solvation shell. 
    
By inspecting insets A and B for the first K-edge peak in Fig.~\ref{Spectra_Cl_LCL_F_FDE}, non-negligible differences in intensity and peak positions among different snapshots are noticeable. These are consequences of the asymmetrical first solvation shell\cite{bouchafra_predictive_2018}, which affects more strongly the lower-lying virtual molecular orbitals of the halide\cite{antalek_solvation_2016}. Therefore, upon taking into account the configurational averaging, we end up with peaks that are broader than those obtained from the individual structures. An opposite behavior is observed in the case of the second peak, which shows no significant changes in intensities and a negligible displacement in peak positions among different snapshots, see Fig.~\ref{sup-Spectra_Cl_ClL_F_sup} in the supplementary material.
    
The \textit{uFDE-rt-TDDFT} results for the {\edgelone}-edge, also shown in Fig.~\ref{Spectra_Cl_LCL_F_FDE}, corresponds to the electronic transition from the 2s~electronic state of chloride anion to~(n+1)p. In contrast to the {\edgek}-edge, there seems to be no significant effect from both the relaxation of the ground-state density and configurational averaging. Thus, the transition to (n+1)p~virtual electronic states presents a ``sharp'' line-like absorption. \cite{wu_x-ray_1995}
    
For the K-edge spectrum of fluoride, shown in Fig.~\ref{Spectra_Cl_LCL_F_FDE}, the \textit{uFDE-rt-TDDFT} paint, in qualitative terms, a very similar picture as for chloride. However, due to the interaction between fluoride and water being significantly stronger than that of chloride (evident from the radial distribution function between fluoride anion and water-oxygen~\cite{real_structural_2016}, which peaks at a distance of \qty{2.6}{\angstrom}, whereas for the average distance between chloride and water-oxygen is almost twice as much at \qty{3.4}{\angstrom} \cite{hofer_solvation_2022}), we see a larger blue shift (\qty{1.09}{\electronvolt}) for the first peak, and a slightly smaller one (\qty{0.13}{\electronvolt}) for the second peak when the ground-state electron density is relaxed (after several FaT cycles). This picture is consistent with the heightened impact of relaxation on valence and low-lying virtual orbitals, and with the suggestion that fluoride forms the most stable hydrate species \cite{hofer_solvation_2022}.
    
Despite the broadening introduced by configurational averaging, the agreement between the \textit{uFDE-rt-TDDFT} chloride K-edge spectra and experiment remains relatively poor. The first feature is not broad enough, which we can attribute to the inability of \textit{uFDE-rt-TDDFT} to reproduce a second peak at \qty{2765}{\electronvolt}, which has been found to arise from the coupling between chloride and the near-neighbor waters~\cite{de_santis_environment_2021}. In addition to that, while the second peak in the \qtyrange{2772}{2775}{\electronvolt} region is present, it is not as broad as in the experiment. 

Determining the {\edgek}-edge XAS spectra of solvated fluoride anion remains a challenge for soft X-ray absorption spectroscopy \cite{smith_soft_2017}. Therefore, direct comparison between our simulated ensemble spectra and experimental data is not possible. Nevertheless, it is instructive to observe the differences and similarities between chloride and fluoride.

\subsubsection{BOMME rt-TDDFT}\label{sec-bomme}

\begin{figure}[!htbp]
   \captionsetup[subfigure]{labelformat=empty}
   \centering
   \begin{subfigure}{0.75\linewidth}
       \includegraphics[width=\linewidth]{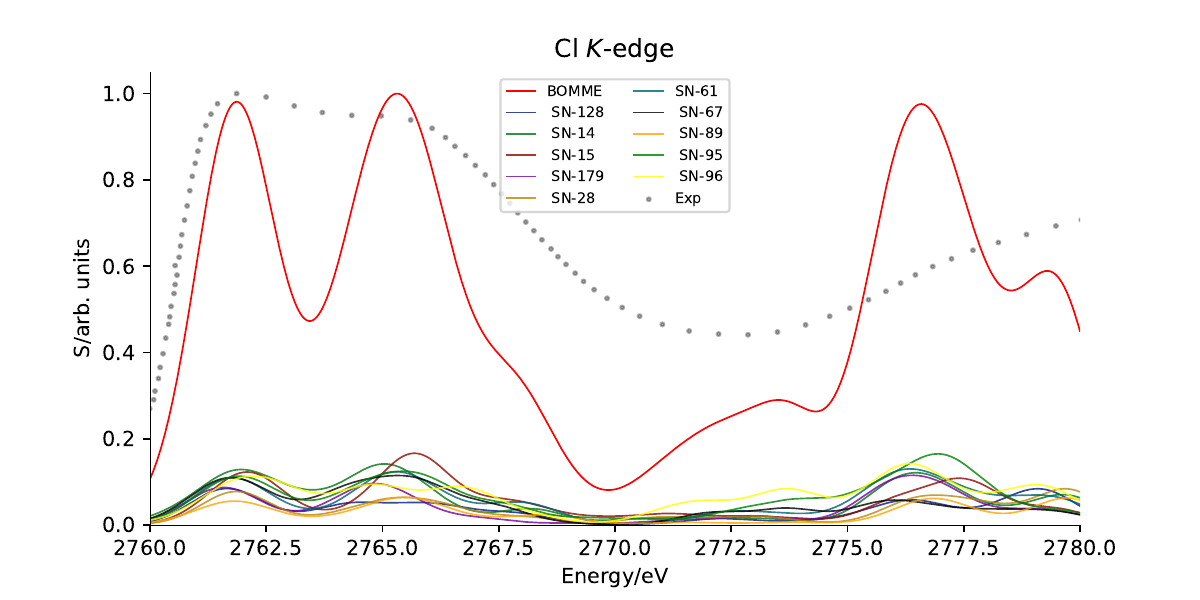}
   \end{subfigure}\hspace{0.1cm}
    \begin{subfigure}{0.75\linewidth}
        \includegraphics[width=\linewidth]{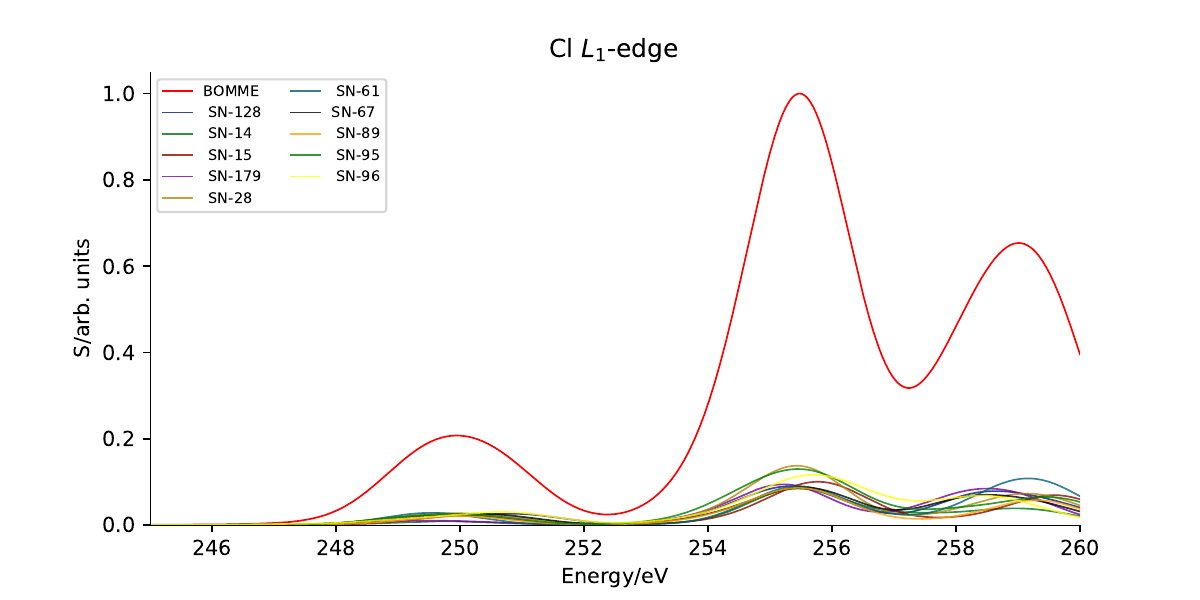}
    \end{subfigure}\hspace{0.1cm}
    \begin{subfigure}{0.75\linewidth}
        \includegraphics[width=\linewidth]{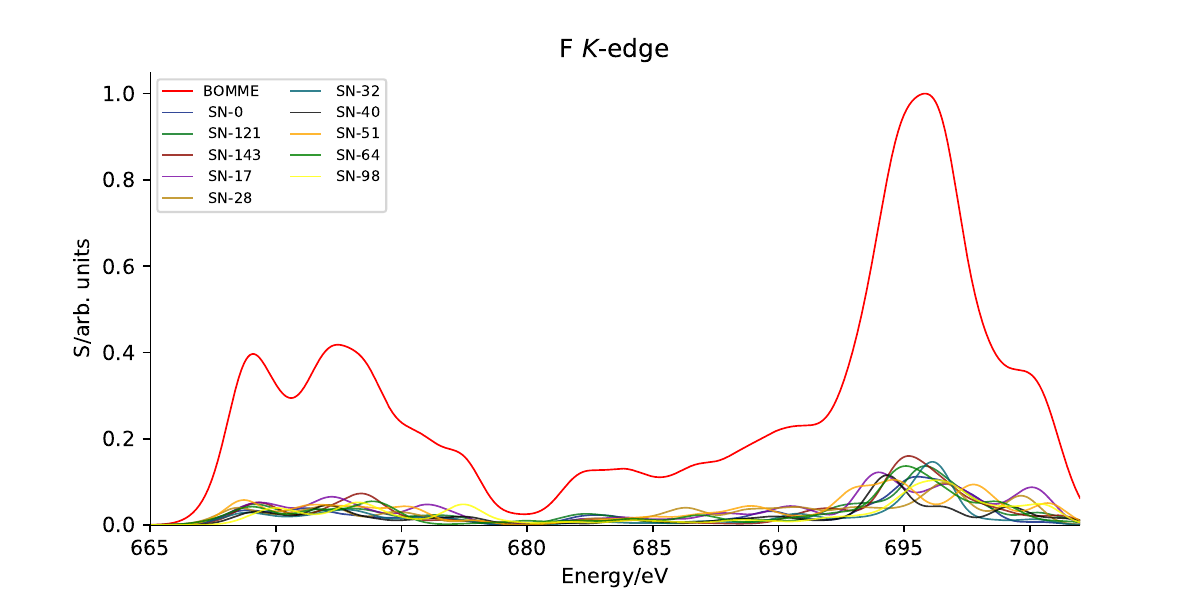}
    \end{subfigure}
    \caption{Chloride {\edgek}-edge (top) and {\edgelone}-edge (middle) and fluoride {\edgek}-edge (bottom) X-ray absorption spectra for the anions embedded in their first solvation shell (\num{8}~water molecules) between the energy range of the free ion edge peak and up to \qty{20}{\electronvolt} (Cl {\edgek}-edge), \qty{14}{\electronvolt} (Cl {\edgelone}-edge), and \qty{35}{\electronvolt} (F {\edgek}-edge). Data was obtained by employing \textit{BOMME-rt-TDDFT}. Insect pictures show the contribution per snapshot. A Gaussian broadening ($\sigma$ = \qty{0.7}{\electronvolt}) was used.
    }
    \label{Spectra_Cl_ClL_F_b}
    \end{figure}

In the top panel of Fig.~\ref{Spectra_Cl_ClL_F_b} we present \textit{BOMME-rt-TDDFT} results for the chloride and fluoride XAS spectra. Starting with the chloride {\edgek}-edge XAS spectra, we see that in contrast to \textit{uFDE-rt-TDDFT}, we have one intense additional peak in the region between \qtyrange{2764}{2768}{\electronvolt}, a less intense one between \qtylist{2770;2772}{\electronvolt}, and a third feature near \qty{2780}{\electronvolt}. We observe that configurational averaging brings about important changes in the spectrum, not only in terms of the broadening of the peaks, but also in terms of intensities, and that different structures contribute differently to different parts of the spectrum; for instance, SN-15 exhibits the highest intensity compared to all other snapshots around the second and third peaks. We examined the radial distribution function (See Fig.~\ref{sup-RDF_Cl_F} in the supplementary information) between chloride and the closest water molecules and observed that the water molecules furthest away from the chloride are closer to it in comparison to other snapshots, explaining the slightly higher intensities.

For the chloride {\edgelone}-edge, as shown in the middle panel of Fig.~\ref{Spectra_Cl_ClL_F_b}, we now observe three peaks: the most intense one in the \qtyrange{254}{256}{\electronvolt} range (as in the \textit{uFDE-rt-TDDFT} results), a significantly less intense peak in the \qtyrange{248}{252}{\electronvolt} range, and another intense peak in the \qtyrange{258}{260}{\electronvolt} range. While the first two peaks were reported before~\cite{de_santis_environment_2021}, the use of configurational averaging not only broadens the less intense peak but also makes it gain in intensity. For the most intense peaks, similar to the {\edgek}-edge, some configurations contribute more than others to the spectrum, while contributions are more homogeneous for the first peak.

In the fluoride {\edgek}-edge spectrum, as seen in the bottom panel of Fig.~\ref{Spectra_Cl_ClL_F_b}, we also observe additional features in the \qtyrange{675}{688}{\electronvolt} region with respect to \textit{uFDE-rt-TDDFT} results, with an overall redshift relative to the latter. However, these are not as intense as the features in the \qtyrange{668}{675}{\electronvolt} and \qtyrange{690}{700}{\electronvolt} regions. A marked difference between fluoride and chloride lies in the contributions from different configurations to the spectrum. 
    
While, as discussed above for chloride, different configurations contribute differently to different regions of the spectrum, for fluoride contributions around \qty{695}{\electronvolt} are significantly more important than those for lower energies. In contrast, for chloride, the contributions are by and large similar across the whole energy range under consideration. This results in a much more intense peak at around \qty{695}{\electronvolt} than in the near-edge region.
    
The additional peaks observed in comparison to \textit{uFDE-rt-TDDFT} were previously attributed to the coupling of the response between chloride and the first solvation shell~\cite{de_santis_environment_2021}. In this analysis, we have further examined the transition density (TD) of a couple of excited states, employing the B3LYP functional for the whole system-- there is a small difference compared to the BOMME calculations, which combine B3LYP for chlorine and BLYP for the water molecules. 
    
The TD analysis indicates (see Fig.~\ref{sup-TD_K_edge_Cl} in the supplementary information) that the first two peaks (within the \qtyrange{2760}{2768}{\electronvolt} window) predominantly arise from 1s~$\rightarrow$~(n+1)p transitions, whereas states in the \qtyrange{2775}{2780}{\electronvolt} correspond to the 1s~$\rightarrow$~(n+2)p 
    transitions, as indicated by their larger spatial extent.
    
The TD for the chloride {\edgelone}-edge (figure \ref{sup-TD_L_edge_Cl} in the supplementary information) and fluoride K-edge paint a similar picture as for the chloride K-edge, with transitions being predominantly centered on the halide. For fluoride (Fig.~\ref{sup-TD_K_edge_F} in the supplementary information), we note distorted 1s~character on the halide in the range of intermediate peaks (\qtyrange{680}{685}{\electronvolt}), while in the final and most prominent peak (\qty{695}{\electronvolt}), we noticed even higher distortions of the halide orbitals, likely caused by the presence of surrounding water molecules. 
    
That said, while the TDs does not show very marked charge-transfer excitations from chloride or fluoride to the water ligands, the comparison between \textit{BOMME-rt-TDDFT} (which allows for contributions from the waters in the excited state) and \textit{uFDE-rt-TDDFT} (which does not) leaves no question as to the importance of going beyond a model in which only the halide belongs to the active subsystem.

Considering now comparison with the experiment for the chloride K-edge, we can observe that the spectral feature in the near-edge (\qtyrange{2760}{2766}{\electronvolt}) is semi-quantitatively recovered in the  \textit{BOMME rt-TDDFT} calculations, with significant differences coming from depletion of intensity around \SI{2763}{\electronvolt}, and a more pronounced decrease in intensity at around \qty{2766}{\electronvolt} in the \textit{BOMME rt-TDDFT} calculations. Furthermore, for energies above \qty{2775}{\electronvolt}, the \textit{BOMME rt-TDDFT} intensities are significantly larger than in the experiment. These represent in any case a significant improvement over the FDE results. 
  
Interestingly, in comparison to the supermolecular result (for a single snapshot, see supplementary information, Fig.~\ref{sup-TD_K_edge_Cl}), we see (a) a better qualitative agreement between \textit{BOMME rt-TDDFT} and experiment for the near-edge feature (below \qty{2770}{\electronvolt}), since the supermolecular calculations show an intense peak at about \qty{2763}{\electronvolt}, with lower-intensity features at \qtylist{2762;2767.5}{\electronvolt}, respectively. On the other hand, supermolecule calculations show less intense features around \qtyrange{2775}{2780}{\electronvolt}. Given our findings from the effect of configurational averaging, we can expect supermolecule calculations to show an overall increase in intensity, but without significant changes in the overall shape of the spectrum.
 
\subsection{\ce{[X(H2O)50]-} models}

Before discussing the \textit{ML-rt-TDDFT} results, we note that we have not carried out \textit{BOMME rt-TDDFT} calculations for the \ce{[X(H2O)50]-} models since the computational cost of such simulations over the different snapshots was beyond the computational means at our disposal. We have on the other hand carried out \textit{uFDE-rt-TDDFT} calculations for these larger systems, with only the halides making up the active subsystem, subsystem 1 and subsystem 2 being the \num{50}-water droplet. However, since we have observed no noticeable difference between these calculations and those discussed above, we do not discuss these further, and the corresponding spectra are shown in Fig.~\ref{sup-Spectra_Cl_50_8}, \ref{sup-FDE_50_8_cl_l}, and \ref{sup-FDE_50_8_f} in the supplementary information.

\subsubsection{Multilevel embedding rt-TDDFT}

It has been shown that the chloride anion significantly changes the structure of water beyond the first solvation shell even outside its second solvation shell \cite{gaiduk_local_2017}.  Thus, in \textit{ML-rt-TDDF} we take into account not only the immediate response of the first solvation shell on the anion, subsystem 1, as discussed in the \textit{\nameref{sec-bomme}} section, but also the extended response of the second and outermost solvation shells, subsystem 2, up to \num{50} water molecules. 
    
\begin{figure}[!htbp]
    \captionsetup[subfigure]{labelformat=empty}
    \centering
    \begin{subfigure}{0.75\linewidth}
        \includegraphics[width=\linewidth]{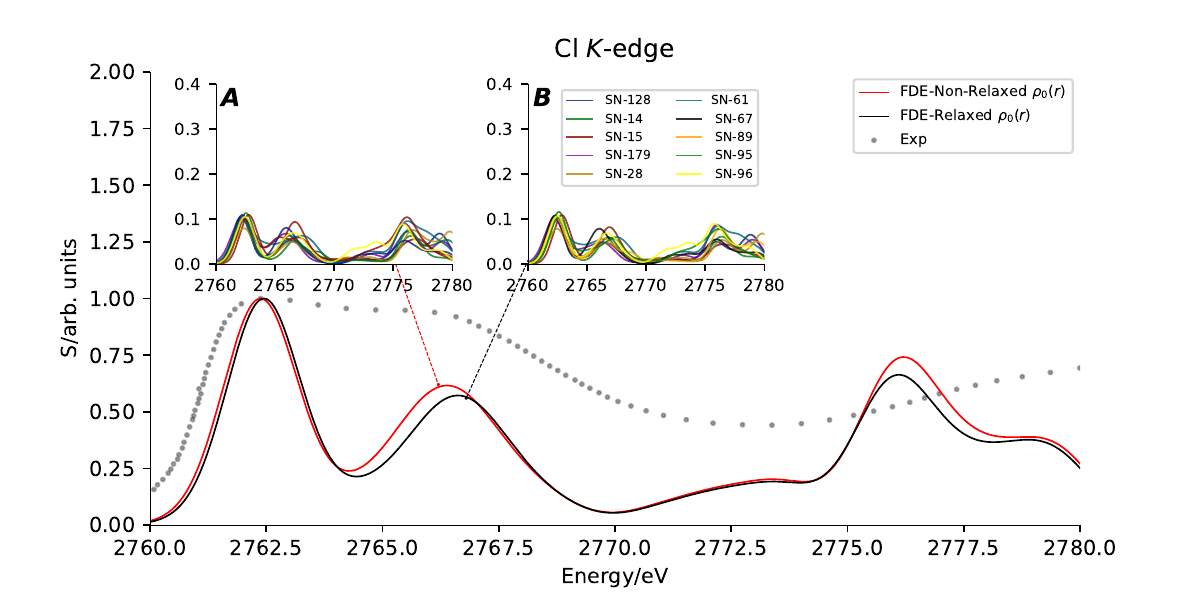}
    \end{subfigure}\hspace{0.1cm}
    \begin{subfigure}{0.75\linewidth}
        \includegraphics[width=\linewidth]{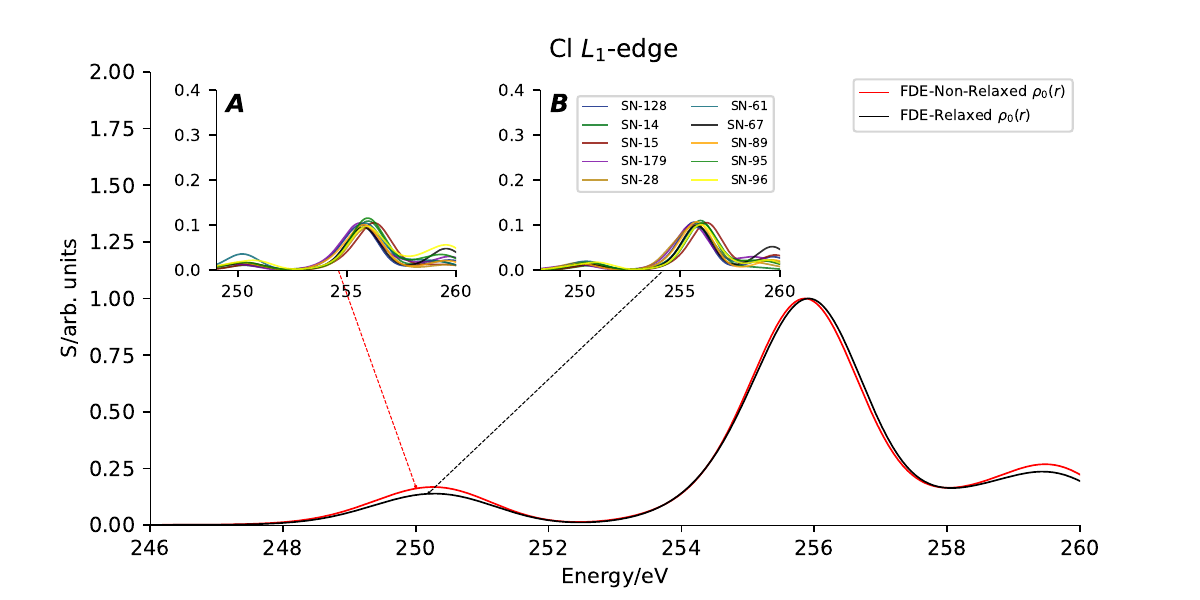}
    \end{subfigure}\hspace{0.1cm}
    \begin{subfigure}{0.75\linewidth}
        \includegraphics[width=\linewidth]{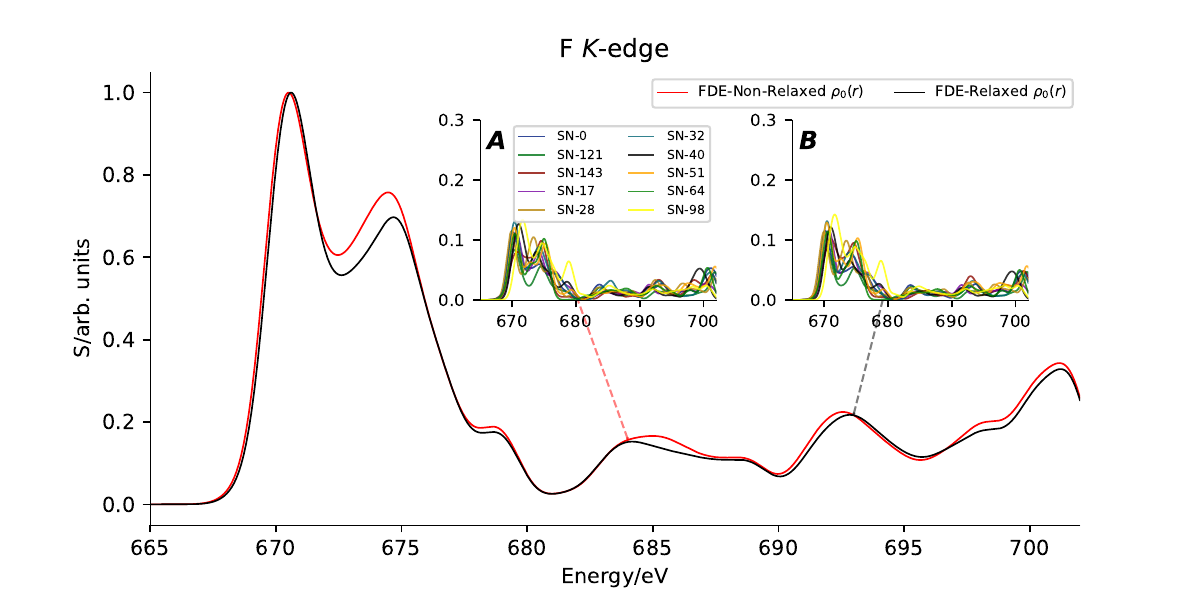}
    \end{subfigure}
    \caption{Chloride \textit{{\edgek}-Edge} (top) and \textit{{\edgelone}-Edge} (middle) and fluoride\textit{{\edgek}-Edge} (bottom) X-ray absorption spectra for the anions embedded in their first solvation shell (8 water molecules) between the energy range of the free ion edge peak and up to \qty{20}{\electronvolt} (Cl {\edgek}-edge), \qty{14}{\electronvolt} (Cl {\edgelone}-edge), and \qty{25}{\electronvolt} (F {\edgek}-edge). Data was obtained by employing \textit{ML-rt-TDDFT} with a non-relax and relaxed initial density $\rho (r)$. 5 cycles of Freeze and Thaw were employed to obtain a relaxed initial density. Insect pictures show the contribution per snapshot. A Gaussian broadening ($\sigma$ = \qty{0.7}{\electronvolt}) was used.
    }
    \label{Spectra_Cl_Cll_F__mlv}
    \end{figure}

The results for the chloride {\edgek}-edge XAS spectra are shown in the top Fig.~\ref{Spectra_Cl_Cll_F__mlv}. Comparing the spectra obtained by \textit{BOMME rt-TDDFT} with those obtained by \textit{ML-rt-TDDF}, see supplementary information Fig. \ref{sup-Spectra_Cl_ClL_F_all}, reveals that in the latter the intensity of features at around \qty{2766}{\electronvolt} and \qty{2776}{\electronvolt} are reduced, while the feature near-edge (\qty{2762.5}{\electronvolt}) remains intense, besides the blue shift of \qty{2.94}{\electronvolt}. With respect to the effect of configurational averaging for the feature at around \qty{2766}{\electronvolt}, we see that contributions from the different snapshots are more similar to each other. We note that from the TD analysis for one snapshot (SN=128, see Fig.~\ref{sup-TD_K_edge_Cl} in the supplementary information), the picture of a mostly chloride-centered, 1s to (n+1)p transition remains for the features below \qty{2770}{\electronvolt}.  
    
With the introduction of a FDE embedding potential in \textit{ML-rt-TDDF}, it becomes of interest to investigate the effect of the relaxation of the ground-state density (after several FaT cycles) for the second solvation shell and beyond has on the spectrum--one would, in any case, expect that this effect to be smaller than for the first solvation shell, since the exact response of the first solvation shell would, by and large, be accounted for by the \textit{BOMME} treatment of the halide subsystem. We confirm this to be the case, with rather small shifts being observed between the  `relaxed'' and ``unrelaxed'' models. However, we note small non-negligible intensities differences between the two models for the features at \qtylist{2766;2776}{\electronvolt}, see top Fig.~\ref{Spectra_Cl_Cll_F__mlv}. 
    
The comparison between \textit{BOMME rt-TDDFT} and \textit{ML-rt-TDDF} chloride {\edgelone}-edge XAS spectrum shows that the latter results in a blue shift of \qty{0.38}{\electronvolt}, see supplementary information Fig.~\ref{sup-Spectra_Cl_ClL_F_all}. The peak with the highest intensity is located at \qty{255.84}{\electronvolt} for \textit{ML-rt-TDDF}. Much like for the K-edge, there are only rather small differences between ``relaxed'' and ``unrelaxed'' models.
    
The difference between \textit{BOMME rt-TDDFT} and \textit{ML-rt-TDDF} for the K-Edge spectrum of fluoride, on the other hand, is remarkable with an energy blue shift of \qty{1.9}{\electronvolt} of \textit{ML-rt-TDDF} with respect to \textit{BOMME rt-TDDFT}, see supplementary information Fig.~\ref{sup-Spectra_Cl_ClL_F_all}. We observe in Fig.~\ref{Spectra_Cl_Cll_F__mlv} that in  \textit{ML-rt-TDDF}, the near-edge peaks are now the most intense ones, with several features within energies above \qty{680}{\electronvolt} with much smaller intensities. While in Fig.~\ref{Spectra_Cl_ClL_F_b}, \textit{BOMME rt-TDDFT} presents the high-intensity feature at \qty{695}{\electronvolt}, this future shows the most significant difference between \textit{BOMME rt-TDDFT} and \textit{ML-rt-TDDF}. In addition to that, the difference between the  `relaxed'' and ``unrelaxed'' models in \textit{ML-rt-TDDF} is significant for the feature at around \qty{675}{\electronvolt}, while for the rest of the spectrum, differences are minor.

Based on our analysis, we can conclude that the intensities of the second feature of the near-edge peak are influenced by the relaxation of the outer shell, subsystem 2, in the simulation of the K-edge for both \ce{Cl-} and \ce{F-} systems. However, the first feature remains unchanged due to its origin from the explicit response of the first solvation shell over the halide, which is treated with \textit{BOMME}. It is this response that will be affected the most by the {\edgek}-edge signal upon density relaxation. Thus, by including a long-range effect through the embedding potential (Eq.~\ref{embpot}), the intensity of the second feature of the near-edge peak decreases around 0.2 arb. units. The overall spectra are in agreement with the spectra obtained using the real-time TDDFT module of NwChem, see supplementary information Fig. \ref{sup-TD_K_edge_F}.

We have successfully addressed two significant challenges that were previously encountered in predicting XAS: (1) the need for a substantial set of solvent configurations to generate accurate XAS spectra. (2) the accuracy of the theory used: particularly the exchange-correlation functional. We have overcome both obstacles by sampling snapshots from a \qty{50}{\nano\second} simulation \cite{bouchafra_predictive_2018} and by using hybrid exchange-correlation functionals to describe the activated subsystem 1 in our $BOMME$ and $ML-rtTDDFT$ approaches\cite{Penfold_200x}.

\section{Conclusions\label{conc}}

In this work, we have presented a novel quantum embedding approach in which projection-based (BOMME) and density-based (FDE) methods are combined. The resulting approach seems particularly well-suited to tackle systems in which some subsystems interact strongly (e.g.\ a solute and its first solvation shell) while others interact more weakly (a solute and its second solvation shell) while maintaining a relatively low computational cost. 

We have applied this approach to investigate the X-ray absorption spectra of chloride and fluoride in aqueous solutions (based on a \num{50}-water droplet model) through real-time time-dependent DFT calculations in order to investigate how core spectra at the K and {\edgelone} edges are changed by employing configurational averaging (by considering several snapshots from classical molecular dynamics simulations with polarizable force fields). Furthermore, we have explored the use of optimized tensor libraries (PyTorch and TensorFlow) to eliminate some of the performance bottlenecks of our prior implementations of real-time TDDFT quantum embedding.

We have found that configuration averaging is important not only as a means of broadening the different features of the spectra but also can have profound effects on the intensities--and in this respect, we have shown that contributions to the final intensities for a particular energy range generally arise from different structures. 

We have also observed that the near-edge features, which predominantly involve the lowest-lying virtual orbitals centered on the halides, are somewhat more sensitive to the quality of the embedding models. This is especially in the case of FDE, for which the representation of charged systems requires the relaxation (several FaT cycles) of ground-state electron density of the first solvation shell, but less so for the other approaches (BOMME) since this relaxation is implicitly taken into account with the optimization of the orbitals for the halide plus first solvation sphere subsystem 1.

Contrary to valence and core ionization energies, for chloride we have found that embedding models in which the halide-water interactions are approximated (e.g.\ by having only the halide in active subsystem region, subsystem 1, in the case of FDE) are not capable of reproducing the experimental spectrum. At the same time, we observed that spectra with the BOMME approach tend to overestimate intensities, especially for features in higher energies. By taking into account the second solvation shell through FDE, this effect is counteracted, resulting in an overall better agreement with the experiment across the energy window considered here.

Finally, given the difficulties associated with the analysis of transitions in real-time TDDFT simulations, for selected structures, we carried out an analysis of transition density for these systems, and from those established that the transitions for the edges under consideration can be considered as dominated by transitions within the halide system. The comparison between embedding approaches shows that, albeit small, the interactions between the halides are nevertheless essential to obtain agreement with experiment results.

\begin{acknowledgement}

MDS, VV, and ASPG acknowledge funding from projects CPER WaveTech, Labex CaPPA (Grant No. ANR-11-LABX-0005-01), CompRIXS (Grant Nos. ANR-19CE29-0019 and DFG JA 2329/6-1), the I-SITE ULNE project OVERSEE and MESONM International Associated Laboratory (LAI) (Grant No. ANR-16-IDEX-0004), as well support from the French regional (Mésocentre de l'Université de Lille) and national supercomputing facilities (Grant Nos. DARI A0110801859 and A0130801859).
JAMB acknowledges support from the Chateaubriand Fellowship of the Office for Science \& Technology of the Embassy of France in the United States. MP acknowledges support from the National Science Foundation under grants No. CHE-2154760 and OAC-2117429.
\end{acknowledgement}

\begin{suppinfo}
The data (input/output) corresponding to the calculations of this paper are available at the Zenodo repository under DOI: {\zenodo}.

The Supporting Information is available free of charge on the \href{http://pubs.acs.org}{ACS Publications website} at DOI: \href{}{XXX}. [add description of what is in the SI].

\end{suppinfo}

\bibliography{refs}

\providecommand{\latin}[1]{#1}
\makeatletter
\providecommand{\doi}
  {\begingroup\let\do\@makeother\dospecials
  \catcode`\{=1 \catcode`\}=2 \doi@aux}
\providecommand{\doi@aux}[1]{\endgroup\texttt{#1}}
\makeatother
\providecommand*\mcitethebibliography{\thebibliography}
\csname @ifundefined\endcsname{endmcitethebibliography}
  {\let\endmcitethebibliography\endthebibliography}{}
\begin{mcitethebibliography}{89}
\providecommand*\natexlab[1]{#1}
\providecommand*\mciteSetBstSublistMode[1]{}
\providecommand*\mciteSetBstMaxWidthForm[2]{}
\providecommand*\mciteBstWouldAddEndPuncttrue
  {\def\EndOfBibitem{\unskip.}}
\providecommand*\mciteBstWouldAddEndPunctfalse
  {\let\EndOfBibitem\relax}
\providecommand*\mciteSetBstMidEndSepPunct[3]{}
\providecommand*\mciteSetBstSublistLabelBeginEnd[3]{}
\providecommand*\EndOfBibitem{}
\mciteSetBstSublistMode{f}
\mciteSetBstMaxWidthForm{subitem}{(\alph{mcitesubitemcount})}
\mciteSetBstSublistLabelBeginEnd
  {\mcitemaxwidthsubitemform\space}
  {\relax}
  {\relax}

\bibitem[Muralidharan \latin{et~al.}(2019)Muralidharan, Pratt, Chaudhari, and
  Rempe]{muralidharan_quasi-chemical_2019}
Muralidharan,~A.; Pratt,~L.; Chaudhari,~M.; Rempe,~S. Quasi-chemical theory for
  anion hydration and specific ion effects: \ce{Cl-}(aq) vs. \ce{F-}(aq).
  \emph{Chem. Phys. Lett.} \textbf{2019}, \emph{737}, 100037, DOI:
  \doi{10.1016/j.cpletx.2019.100037}\relax
\mciteBstWouldAddEndPuncttrue
\mciteSetBstMidEndSepPunct{\mcitedefaultmidpunct}
{\mcitedefaultendpunct}{\mcitedefaultseppunct}\relax
\EndOfBibitem
\bibitem[Hofer(2022)]{hofer_solvation_2022}
Hofer,~T.~S. Solvation {Structure} and {Ion}–{Solvent} {Hydrogen} {Bonding}
  of {Hydrated} {Fluoride}, {Chloride} and {Bromide}—{A} {Comparative}
  {QM}/{MM} {MD} {Simulation} {Study}. \emph{Liquids} \textbf{2022}, \emph{2},
  445--464, DOI: \doi{10.3390/liquids2040026}\relax
\mciteBstWouldAddEndPuncttrue
\mciteSetBstMidEndSepPunct{\mcitedefaultmidpunct}
{\mcitedefaultendpunct}{\mcitedefaultseppunct}\relax
\EndOfBibitem
\bibitem[Chowdhuri and Chandra(2006)Chowdhuri, and
  Chandra]{chowdhuri_dynamics_2006}
Chowdhuri,~S.; Chandra,~A. Dynamics of {Halide} {Ion}-{Water} {Hydrogen}
  {Bonds} in {Aqueous} {Solutions}: {Dependence} on {Ion} {Size} and
  {Temperature}. \emph{J. Phys. Chem. B} \textbf{2006}, \emph{110}, 9674--9680,
  DOI: \doi{10.1021/jp057544d}\relax
\mciteBstWouldAddEndPuncttrue
\mciteSetBstMidEndSepPunct{\mcitedefaultmidpunct}
{\mcitedefaultendpunct}{\mcitedefaultseppunct}\relax
\EndOfBibitem
\bibitem[Heuft and Meijer(2005)Heuft, and Meijer]{heuft_density_2005}
Heuft,~J.~M.; Meijer,~E.~J. Density functional theory based molecular-dynamics
  study of aqueous fluoride solvation. \emph{J. Chem. Phys.} \textbf{2005},
  \emph{122}, 094501, DOI: \doi{10.1063/1.1853352}\relax
\mciteBstWouldAddEndPuncttrue
\mciteSetBstMidEndSepPunct{\mcitedefaultmidpunct}
{\mcitedefaultendpunct}{\mcitedefaultseppunct}\relax
\EndOfBibitem
\bibitem[Bakker(2008)]{bakker_structural_2008}
Bakker,~H.~J. Structural {Dynamics} of {Aqueous} {Salt} {Solutions}.
  \emph{Chem. Rev.} \textbf{2008}, \emph{108}, 1456--1473, DOI:
  \doi{10.1021/cr0206622}\relax
\mciteBstWouldAddEndPuncttrue
\mciteSetBstMidEndSepPunct{\mcitedefaultmidpunct}
{\mcitedefaultendpunct}{\mcitedefaultseppunct}\relax
\EndOfBibitem
\bibitem[Réal \latin{et~al.}(2016)Réal, Severo Pereira~Gomes,
  Guerrero~Martínez, Ayed, Galland, Masella, and Vallet]{real_structural_2016}
Réal,~F.; Severo Pereira~Gomes,~A.; Guerrero~Martínez,~Y.~O.; Ayed,~T.;
  Galland,~N.; Masella,~M.; Vallet,~V. Structural, dynamical, and transport
  properties of the hydrated halides: {How} do {At} $^{\textrm{-}}$ bulk
  properties compare with those of the other halides, from {F} $^{\textrm{-}}$
  to {I} $^{\textrm{-}}$ ? \emph{J. Chem. Phys.} \textbf{2016}, \emph{144},
  124513, DOI: \doi{10.1063/1.4944613}\relax
\mciteBstWouldAddEndPuncttrue
\mciteSetBstMidEndSepPunct{\mcitedefaultmidpunct}
{\mcitedefaultendpunct}{\mcitedefaultseppunct}\relax
\EndOfBibitem
\bibitem[Antalek \latin{et~al.}(2016)Antalek, Pace, Hedman, Hodgson, Chillemi,
  Benfatto, Sarangi, and Frank]{antalek_solvation_2016}
Antalek,~M.; Pace,~E.; Hedman,~B.; Hodgson,~K.~O.; Chillemi,~G.; Benfatto,~M.;
  Sarangi,~R.; Frank,~P. Solvation structure of the halides from x-ray
  absorption spectroscopy. \emph{J. Chem. Phys.} \textbf{2016}, \emph{145},
  044318, DOI: \doi{10.1063/1.4959589}\relax
\mciteBstWouldAddEndPuncttrue
\mciteSetBstMidEndSepPunct{\mcitedefaultmidpunct}
{\mcitedefaultendpunct}{\mcitedefaultseppunct}\relax
\EndOfBibitem
\bibitem[Marin \latin{et~al.}(2021)Marin, Janik, Bartels, and
  Chipman]{marin_failure_2021}
Marin,~T.~W.; Janik,~I.; Bartels,~D.~M.; Chipman,~D.~M. Failure of molecular
  dynamics to provide appropriate structures for quantum mechanical description
  of the aqueous chloride ion charge-transfer-to-solvent ultraviolet spectrum.
  \emph{Phys. Chem. Chem. Phys.} \textbf{2021}, \emph{23}, 9109--9120, DOI:
  \doi{10.1039/D1CP00930C}\relax
\mciteBstWouldAddEndPuncttrue
\mciteSetBstMidEndSepPunct{\mcitedefaultmidpunct}
{\mcitedefaultendpunct}{\mcitedefaultseppunct}\relax
\EndOfBibitem
\bibitem[Waluyo \latin{et~al.}(2014)Waluyo, Nordlund, Bergmann, Schlesinger,
  Pettersson, and Nilsson]{waluyo_different_2014}
Waluyo,~I.; Nordlund,~D.; Bergmann,~U.; Schlesinger,~D.; Pettersson,~L. G.~M.;
  Nilsson,~A. A different view of structure-making and structure-breaking in
  alkali halide aqueous solutions through x-ray absorption spectroscopy.
  \emph{J. Chem. Phys.} \textbf{2014}, \emph{140}, 244506, DOI:
  \doi{10.1063/1.4881600}\relax
\mciteBstWouldAddEndPuncttrue
\mciteSetBstMidEndSepPunct{\mcitedefaultmidpunct}
{\mcitedefaultendpunct}{\mcitedefaultseppunct}\relax
\EndOfBibitem
\bibitem[Collins(2019)]{collins_behavior_2019}
Collins,~K.~D. The behavior of ions in water is controlled by their water
  affinity. \emph{Q. Rev. Biophys.} \textbf{2019}, \emph{52}, e11, DOI:
  \doi{10.1017/S0033583519000106}\relax
\mciteBstWouldAddEndPuncttrue
\mciteSetBstMidEndSepPunct{\mcitedefaultmidpunct}
{\mcitedefaultendpunct}{\mcitedefaultseppunct}\relax
\EndOfBibitem
\bibitem[Fang \latin{et~al.}(2023)Fang, Li, Wang, Duan, and
  Zhao]{fang_long-life_2023}
Fang,~Z.; Li,~M.; Wang,~L.; Duan,~X.; Zhao,~H. A long-life aqueous
  {Fluoride}-ion battery based on {Water}-in-salt electrolyte. \emph{Inorg.
  Chem. Commun.} \textbf{2023}, \emph{148}, 110275, DOI:
  \doi{10.1016/j.inoche.2022.110275}\relax
\mciteBstWouldAddEndPuncttrue
\mciteSetBstMidEndSepPunct{\mcitedefaultmidpunct}
{\mcitedefaultendpunct}{\mcitedefaultseppunct}\relax
\EndOfBibitem
\bibitem[Mao \latin{et~al.}(2022)Mao, Zhou, Xu, Wu, Wang, Xiong, Wu, and
  Shi]{mao_molecular_2022}
Mao,~Y.; Zhou,~T.; Xu,~L.; Wu,~W.; Wang,~R.; Xiong,~Z.; Wu,~D.; Shi,~H.
  Molecular understanding of aqueous electrolyte properties and dielectric
  effect in a {CDI} system. \emph{Chem. Eng. J.} \textbf{2022}, \emph{435},
  134750, DOI: \doi{10.1016/j.cej.2022.134750}\relax
\mciteBstWouldAddEndPuncttrue
\mciteSetBstMidEndSepPunct{\mcitedefaultmidpunct}
{\mcitedefaultendpunct}{\mcitedefaultseppunct}\relax
\EndOfBibitem
\bibitem[Kulik \latin{et~al.}(2010)Kulik, Marzari, Correa, Prendergast,
  Schwegler, and Galli]{kulik_local_2010}
Kulik,~H.~J.; Marzari,~N.; Correa,~A.~A.; Prendergast,~D.; Schwegler,~E.;
  Galli,~G. Local Effects in the X-ray Absorption Spectrum of Salt Water.
  \emph{J. Phys. Chem. B} \textbf{2010}, \emph{114}, 9594--9601, DOI:
  \doi{10.1021/jp103526y}\relax
\mciteBstWouldAddEndPuncttrue
\mciteSetBstMidEndSepPunct{\mcitedefaultmidpunct}
{\mcitedefaultendpunct}{\mcitedefaultseppunct}\relax
\EndOfBibitem
\bibitem[Alayoglu \latin{et~al.}(2023)Alayoglu, Chang, Yan, Chen, and
  Mankad]{alayoglu_uncovering_2023}
Alayoglu,~P.; Chang,~T.; Yan,~C.; Chen,~Y.; Mankad,~N.~P. Uncovering a \ce{CF3}
  {Effect} on {X}‐ray {Absorption} {Energies} of \ce{[Cu(CF3)4]-} and
  {Related} {Copper} {Compounds} by {Using} {Resonant} {Diffraction}
  {Anomalous} {Fine} {Structure} ({DAFS}) {Measurements}**. \emph{Angew. Chem.}
  \textbf{2023}, e202313744, DOI: \doi{10.1002/ange.202313744}\relax
\mciteBstWouldAddEndPuncttrue
\mciteSetBstMidEndSepPunct{\mcitedefaultmidpunct}
{\mcitedefaultendpunct}{\mcitedefaultseppunct}\relax
\EndOfBibitem
\bibitem[Zhang \latin{et~al.}(2023)Zhang, Xie, Long, Günzing, Wende, Ollefs,
  and Zhang]{zhang_autonomous_2023}
Zhang,~Y.; Xie,~R.; Long,~T.; Günzing,~D.; Wende,~H.; Ollefs,~K.~J.; Zhang,~H.
  Autonomous atomic {Hamiltonian} construction and active sampling of {X}-ray
  absorption spectroscopy by adversarial {Bayesian} optimization. \emph{npj
  Comput. Mater.} \textbf{2023}, \emph{9}, 46, DOI:
  \doi{10.1038/s41524-023-00994-w}\relax
\mciteBstWouldAddEndPuncttrue
\mciteSetBstMidEndSepPunct{\mcitedefaultmidpunct}
{\mcitedefaultendpunct}{\mcitedefaultseppunct}\relax
\EndOfBibitem
\bibitem[Zhao \latin{et~al.}(2007)Zhao, Chu, Yang, Yu, Chen, Guo, Zhou, Shi,
  Marcelli, Niu, Teng, Gong, Benfatto, and Wu]{zhao_quantitative_2007}
Zhao,~W.; Chu,~W.; Yang,~F.; Yu,~M.; Chen,~D.; Guo,~X.; Zhou,~D.; Shi,~N.;
  Marcelli,~A.; Niu,~L.; Teng,~M.; Gong,~W.; Benfatto,~M.; Wu,~Z. Quantitative
  investigation of two metallohydrolases by {X}-ray absorption spectroscopy
  near-edge spectroscopy. \emph{Nucl. Instrum. Methods Phys. Res., Sect. A}
  \textbf{2007}, \emph{580}, 451--456, DOI:
  \doi{10.1016/j.nima.2007.05.196}\relax
\mciteBstWouldAddEndPuncttrue
\mciteSetBstMidEndSepPunct{\mcitedefaultmidpunct}
{\mcitedefaultendpunct}{\mcitedefaultseppunct}\relax
\EndOfBibitem
\bibitem[Tofoni \latin{et~al.}(2023)Tofoni, Tavani, Persson, and
  D’Angelo]{tofoni_p_2023}
Tofoni,~A.; Tavani,~F.; Persson,~I.; D’Angelo,~P. P {K}-{Edge} {XANES}
  {Calculations} of {Mineral} {Standards}: {Exploring} the {Potential} of
  {Theoretical} {Methods} in the {Analysis} of {Phosphorus} {Speciation}.
  \emph{Inorg. Chem.} \textbf{2023}, \emph{62}, 11188--11198, DOI:
  \doi{10.1021/acs.inorgchem.3c01346}\relax
\mciteBstWouldAddEndPuncttrue
\mciteSetBstMidEndSepPunct{\mcitedefaultmidpunct}
{\mcitedefaultendpunct}{\mcitedefaultseppunct}\relax
\EndOfBibitem
\bibitem[Herbert \latin{et~al.}(2023)Herbert, Zhu, Alam, and
  Ojha]{herbert_time-dependent_2023}
Herbert,~J.~M.; Zhu,~Y.; Alam,~B.; Ojha,~A.~K. Time-{Dependent} {Density}
  {Functional} {Theory} for {X}-ray {Absorption} {Spectra}: {Comparing} the
  {Real}-{Time} {Approach} to {Linear} {Response}. \emph{J. Chem. Theory
  Comput.} \textbf{2023}, \emph{19}, 6745--6760, DOI:
  \doi{10.1021/acs.jctc.3c00673}\relax
\mciteBstWouldAddEndPuncttrue
\mciteSetBstMidEndSepPunct{\mcitedefaultmidpunct}
{\mcitedefaultendpunct}{\mcitedefaultseppunct}\relax
\EndOfBibitem
\bibitem[Misael and Severo Pereira~Gomes(2023)Misael, and Severo
  Pereira~Gomes]{misael_core_2023}
Misael,~W.~A.; Severo Pereira~Gomes,~A. Core {Excitations} of {Uranyl} in
  \ce{Cs2UO2Cl4} from {Relativistic} {Embedded} {Damped} {Response}
  {Time}-{Dependent} {Density} {Functional} {Theory} {Calculations}.
  \emph{Inorg. Chem.} \textbf{2023}, \emph{62}, 11589--11601, DOI:
  \doi{10.1021/acs.inorgchem.3c01302}\relax
\mciteBstWouldAddEndPuncttrue
\mciteSetBstMidEndSepPunct{\mcitedefaultmidpunct}
{\mcitedefaultendpunct}{\mcitedefaultseppunct}\relax
\EndOfBibitem
\bibitem[Park \latin{et~al.}(2023)Park, Komarov, Lee, and
  Choi]{park_mixed-reference_2023}
Park,~W.; Komarov,~K.; Lee,~S.; Choi,~C.~H. Mixed-{Reference} {Spin}-{Flip}
  {Time}-{Dependent} {Density} {Functional} {Theory}: {Multireference}
  {Advantages} with the {Practicality} of {Linear} {Response} {Theory}.
  \emph{J. Phys. Chem. Lett.} \textbf{2023}, \emph{14}, 8896--8908, DOI:
  \doi{10.1021/acs.jpclett.3c02296}\relax
\mciteBstWouldAddEndPuncttrue
\mciteSetBstMidEndSepPunct{\mcitedefaultmidpunct}
{\mcitedefaultendpunct}{\mcitedefaultseppunct}\relax
\EndOfBibitem
\bibitem[Fransson and Pettersson(2023)Fransson, and
  Pettersson]{fransson_calibrating_2023}
Fransson,~T.; Pettersson,~L. G.~M. Calibrating {TDDFT} {Calculations} of the
  {X}-ray {Emission} {Spectrum} of {Liquid} {Water}: {The} {Effects} of
  {Hartree}–{Fock} {Exchange}. \emph{J. Chem. Theory Comput.} \textbf{2023},
  \emph{19}, 7333--7342, DOI: \doi{10.1021/acs.jctc.3c00728}\relax
\mciteBstWouldAddEndPuncttrue
\mciteSetBstMidEndSepPunct{\mcitedefaultmidpunct}
{\mcitedefaultendpunct}{\mcitedefaultseppunct}\relax
\EndOfBibitem
\bibitem[Liekhus-Schmaltz \latin{et~al.}(2021)Liekhus-Schmaltz, Ho, Weakly,
  Aquila, Schoenlein, Khalil, and Govind]{liekhus-schmaltz_ultrafast_2021}
Liekhus-Schmaltz,~C.~E.; Ho,~P.~J.; Weakly,~R.~B.; Aquila,~A.;
  Schoenlein,~R.~W.; Khalil,~M.; Govind,~N. Ultrafast x-ray pump x-ray probe
  transient absorption spectroscopy: {A} computational study and proposed
  experiment probing core-valence electronic correlations in solvated
  complexes. \emph{J. Chem. Phys.} \textbf{2021}, \emph{154}, 214107, DOI:
  \doi{10.1063/5.0047381}\relax
\mciteBstWouldAddEndPuncttrue
\mciteSetBstMidEndSepPunct{\mcitedefaultmidpunct}
{\mcitedefaultendpunct}{\mcitedefaultseppunct}\relax
\EndOfBibitem
\bibitem[Pezeshki and Lin(2014)Pezeshki, and Lin]{pezeshki_molecular_2014}
Pezeshki,~S.; Lin,~H. Molecular dynamics simulations of ion solvation by
  flexible-boundary {QM}/{MM}: {On}-the-fly partial charge transfer between
  {QM} and {MM} subsystems. \emph{J. Comput. Chem.} \textbf{2014}, \emph{35},
  1778--1788, DOI: \doi{10.1002/jcc.23685}\relax
\mciteBstWouldAddEndPuncttrue
\mciteSetBstMidEndSepPunct{\mcitedefaultmidpunct}
{\mcitedefaultendpunct}{\mcitedefaultseppunct}\relax
\EndOfBibitem
\bibitem[Gomez \latin{et~al.}(2022)Gomez, Pratt, Asthagiri, and
  Rempe]{gomez_hydrated_2022}
Gomez,~D.~T.; Pratt,~L.~R.; Asthagiri,~D.~N.; Rempe,~S.~B. Hydrated {Anions}:
  {From} {Clusters} to {Bulk} {Solution} with {Quasi}-{Chemical} {Theory}.
  \emph{Acc. Chem. Res.} \textbf{2022}, \emph{55}, 2201--2212, DOI:
  \doi{10.1021/acs.accounts.2c00078}\relax
\mciteBstWouldAddEndPuncttrue
\mciteSetBstMidEndSepPunct{\mcitedefaultmidpunct}
{\mcitedefaultendpunct}{\mcitedefaultseppunct}\relax
\EndOfBibitem
\bibitem[Andreussi and Fisicaro(2019)Andreussi, and
  Fisicaro]{andreussi_continuum_2019}
Andreussi,~O.; Fisicaro,~G. Continuum embeddings in condensed‐matter
  simulations. \emph{Int. J. Quantum Chem.} \textbf{2019}, \emph{119}, e25725,
  DOI: \doi{10.1002/qua.25725}\relax
\mciteBstWouldAddEndPuncttrue
\mciteSetBstMidEndSepPunct{\mcitedefaultmidpunct}
{\mcitedefaultendpunct}{\mcitedefaultseppunct}\relax
\EndOfBibitem
\bibitem[Milanese \latin{et~al.}(2017)Milanese, Provorse, Alameda, and
  Isborn]{milanese_convergence_2017}
Milanese,~J.~M.; Provorse,~M.~R.; Alameda,~E.; Isborn,~C.~M. Convergence of
  {Computed} {Aqueous} {Absorption} {Spectra} with {Explicit} {Quantum}
  {Mechanical} {Solvent}. \emph{J. Chem. Theory Comput.} \textbf{2017},
  \emph{13}, 2159--2171, DOI: \doi{10.1021/acs.jctc.7b00159}\relax
\mciteBstWouldAddEndPuncttrue
\mciteSetBstMidEndSepPunct{\mcitedefaultmidpunct}
{\mcitedefaultendpunct}{\mcitedefaultseppunct}\relax
\EndOfBibitem
\bibitem[Ding \latin{et~al.}(2017)Ding, Manby, and Miller]{ding_embedded_2017}
Ding,~F.; Manby,~F.~R.; Miller,~T.~F. Embedded {Mean}-{Field} {Theory} with
  {Block}-{Orthogonalized} {Partitioning}. \emph{J. Chem. Theory Comput.}
  \textbf{2017}, \emph{13}, 1605--1615, DOI:
  \doi{10.1021/acs.jctc.6b01065}\relax
\mciteBstWouldAddEndPuncttrue
\mciteSetBstMidEndSepPunct{\mcitedefaultmidpunct}
{\mcitedefaultendpunct}{\mcitedefaultseppunct}\relax
\EndOfBibitem
\bibitem[Sharma and Sierka(2022)Sharma, and Sierka]{sharma_efficient_2022}
Sharma,~M.; Sierka,~M. Efficient {Implementation} of {Density} {Functional}
  {Theory} {Based} {Embedding} for {Molecular} and {Periodic} {Systems} {Using}
  {Gaussian} {Basis} {Functions}. \emph{J. Chem. Theory Comput.} \textbf{2022},
  \emph{18}, 6892--6904, DOI: \doi{10.1021/acs.jctc.2c00380}\relax
\mciteBstWouldAddEndPuncttrue
\mciteSetBstMidEndSepPunct{\mcitedefaultmidpunct}
{\mcitedefaultendpunct}{\mcitedefaultseppunct}\relax
\EndOfBibitem
\bibitem[Wesolowski \latin{et~al.}(2015)Wesolowski, Shedge, and
  Zhou]{wesolowski_frozen-density_2015}
Wesolowski,~T.~A.; Shedge,~S.; Zhou,~X. Frozen-{Density} {Embedding} {Strategy}
  for {Multilevel} {Simulations} of {Electronic} {Structure}. \emph{Chem. Rev.}
  \textbf{2015}, \emph{115}, 5891--5928, DOI: \doi{10.1021/cr500502v}\relax
\mciteBstWouldAddEndPuncttrue
\mciteSetBstMidEndSepPunct{\mcitedefaultmidpunct}
{\mcitedefaultendpunct}{\mcitedefaultseppunct}\relax
\EndOfBibitem
\bibitem[Iannuzzi \latin{et~al.}(2006)Iannuzzi, Kirchner, and
  Hutter]{iannuzzi_density_2006}
Iannuzzi,~M.; Kirchner,~B.; Hutter,~J. Density functional embedding for
  molecular systems. \emph{Chem. Phys. Lett.} \textbf{2006}, \emph{421},
  16--20, DOI: \doi{10.1016/j.cplett.2005.08.155}\relax
\mciteBstWouldAddEndPuncttrue
\mciteSetBstMidEndSepPunct{\mcitedefaultmidpunct}
{\mcitedefaultendpunct}{\mcitedefaultseppunct}\relax
\EndOfBibitem
\bibitem[Krishtal \latin{et~al.}(2015)Krishtal, Ceresoli, and
  Pavanello]{krishtal_subsystem_2015}
Krishtal,~A.; Ceresoli,~D.; Pavanello,~M. Subsystem real-time time dependent
  density functional theory. \emph{J. Chem. Phys.} \textbf{2015}, \emph{142},
  154116, DOI: \doi{10.1063/1.4918276}\relax
\mciteBstWouldAddEndPuncttrue
\mciteSetBstMidEndSepPunct{\mcitedefaultmidpunct}
{\mcitedefaultendpunct}{\mcitedefaultseppunct}\relax
\EndOfBibitem
\bibitem[Jacob and Neugebauer(2014)Jacob, and Neugebauer]{jacob_subsystem_2014}
Jacob,~C.~R.; Neugebauer,~J. Subsystem density-functional theory: {Subsystem}
  density-functional theory. \emph{Wiley Interdiscip. Rev.: Comput. Mol. Sci.}
  \textbf{2014}, \emph{4}, 325--362, DOI: \doi{10.1002/wcms.1175}\relax
\mciteBstWouldAddEndPuncttrue
\mciteSetBstMidEndSepPunct{\mcitedefaultmidpunct}
{\mcitedefaultendpunct}{\mcitedefaultseppunct}\relax
\EndOfBibitem
\bibitem[Fu and Wesolowski(2023)Fu, and Wesolowski]{fu_excitation_2023}
Fu,~M.; Wesolowski,~T.~A. Excitation {Energies} of {Embedded} {Chromophores}
  from {Frozen}-{Density} {Embedding} {Theory} {Using} {State}-{Specific}
  {Electron} {Densities} of the {Environment}. \emph{J. Phys. Chem. A}
  \textbf{2023}, \emph{127}, 535--545, DOI:
  \doi{10.1021/acs.jpca.2c07747}\relax
\mciteBstWouldAddEndPuncttrue
\mciteSetBstMidEndSepPunct{\mcitedefaultmidpunct}
{\mcitedefaultendpunct}{\mcitedefaultseppunct}\relax
\EndOfBibitem
\bibitem[Niemeyer \latin{et~al.}(2023)Niemeyer, Eschenbach, Bensberg, Tölle,
  Hellmann, Lampe, Massolle, Rikus, Schnieders, Unsleber, and
  Neugebauer]{niemeyer_subsystem_2023}
Niemeyer,~N.; Eschenbach,~P.; Bensberg,~M.; Tölle,~J.; Hellmann,~L.;
  Lampe,~L.; Massolle,~A.; Rikus,~A.; Schnieders,~D.; Unsleber,~J.~P.;
  Neugebauer,~J. The subsystem quantum chemistry program \textsc{Serenity}.
  \emph{WIREs Comput. Mol. Sci.} \textbf{2023}, \emph{13}, e1647, DOI:
  \doi{10.1002/wcms.1647}\relax
\mciteBstWouldAddEndPuncttrue
\mciteSetBstMidEndSepPunct{\mcitedefaultmidpunct}
{\mcitedefaultendpunct}{\mcitedefaultseppunct}\relax
\EndOfBibitem
\bibitem[Sen and Visscher(2023)Sen, and Visscher]{sen_towards_2023}
Sen,~S.; Visscher,~L. Towards the description of charge transfer states in
  solubilised {LHCII} using subsystem {DFT}. \emph{Photosynth. Res.}
  \textbf{2023}, \emph{156}, 39--57, DOI:
  \doi{10.1007/s11120-022-00950-7}\relax
\mciteBstWouldAddEndPuncttrue
\mciteSetBstMidEndSepPunct{\mcitedefaultmidpunct}
{\mcitedefaultendpunct}{\mcitedefaultseppunct}\relax
\EndOfBibitem
\bibitem[Bouchafra \latin{et~al.}(2018)Bouchafra, Shee, Réal, Vallet, and
  Severo Pereira~Gomes]{bouchafra_predictive_2018}
Bouchafra,~Y.; Shee,~A.; Réal,~F.; Vallet,~V.; Severo Pereira~Gomes,~A.
  Predictive {Simulations} of {Ionization} {Energies} of {Solvated} {Halide}
  {Ions} with {Relativistic} {Embedded} {Equation} of {Motion} {Coupled}
  {Cluster} {Theory}. \emph{Phys. Rev. Lett.} \textbf{2018}, \emph{121},
  266001, DOI: \doi{10.1103/PhysRevLett.121.266001}\relax
\mciteBstWouldAddEndPuncttrue
\mciteSetBstMidEndSepPunct{\mcitedefaultmidpunct}
{\mcitedefaultendpunct}{\mcitedefaultseppunct}\relax
\EndOfBibitem
\bibitem[Opoku \latin{et~al.}(2022)Opoku, Toubin, and
  Gomes]{opoku_simulating_2022}
Opoku,~R.~A.; Toubin,~C.; Gomes,~A. S.~P. Simulating core electron binding
  energies of halogenated species adsorbed on ice surfaces and in solution
  \textit{via} relativistic quantum embedding calculations. \emph{Phys. Chem.
  Chem. Phys.} \textbf{2022}, \emph{24}, 14390--14407, DOI:
  \doi{10.1039/D1CP05836C}\relax
\mciteBstWouldAddEndPuncttrue
\mciteSetBstMidEndSepPunct{\mcitedefaultmidpunct}
{\mcitedefaultendpunct}{\mcitedefaultseppunct}\relax
\EndOfBibitem
\bibitem[Höfener \latin{et~al.}(2013)Höfener, Gomes, and
  Visscher]{hofener_solvatochromic_2013}
Höfener,~S.; Gomes,~A. S.~P.; Visscher,~L. Solvatochromic shifts from
  coupled-cluster theory embedded in density functional theory. \emph{J. Chem.
  Phys.} \textbf{2013}, \emph{139}, 104106, DOI: \doi{10.1063/1.4820488}\relax
\mciteBstWouldAddEndPuncttrue
\mciteSetBstMidEndSepPunct{\mcitedefaultmidpunct}
{\mcitedefaultendpunct}{\mcitedefaultseppunct}\relax
\EndOfBibitem
\bibitem[De~Santis \latin{et~al.}(2020)De~Santis, Belpassi, Jacob, Severo
  Pereira~Gomes, Tarantelli, Visscher, and
  Storchi]{de_santis_environmental_2020}
De~Santis,~M.; Belpassi,~L.; Jacob,~C.~R.; Severo Pereira~Gomes,~A.;
  Tarantelli,~F.; Visscher,~L.; Storchi,~L. Environmental {Effects} with
  {Frozen}-{Density} {Embedding} in {Real}-{Time} {Time}-{Dependent} {Density}
  {Functional} {Theory} {Using} {Localized} {Basis} {Functions}. \emph{J. Chem.
  Theory Comput.} \textbf{2020}, \emph{16}, 5695--5711, DOI:
  \doi{10.1021/acs.jctc.0c00603}\relax
\mciteBstWouldAddEndPuncttrue
\mciteSetBstMidEndSepPunct{\mcitedefaultmidpunct}
{\mcitedefaultendpunct}{\mcitedefaultseppunct}\relax
\EndOfBibitem
\bibitem[De~Santis \latin{et~al.}(2022)De~Santis, Sorbelli, Vallet, Gomes,
  Storchi, and Belpassi]{de_santis_frozen-density_2022}
De~Santis,~M.; Sorbelli,~D.; Vallet,~V.; Gomes,~A. S.~P.; Storchi,~L.;
  Belpassi,~L. Frozen-{Density} {Embedding} for {Including} {Environmental}
  {Effects} in the {Dirac}-{Kohn}–{Sham} {Theory}: {An} {Implementation}
  {Based} on {Density} {Fitting} and {Prototyping} {Techniques}. \emph{J. Chem.
  Theory Comput.} \textbf{2022}, \emph{18}, 5992--6009, DOI:
  \doi{10.1021/acs.jctc.2c00499}\relax
\mciteBstWouldAddEndPuncttrue
\mciteSetBstMidEndSepPunct{\mcitedefaultmidpunct}
{\mcitedefaultendpunct}{\mcitedefaultseppunct}\relax
\EndOfBibitem
\bibitem[De~Santis \latin{et~al.}(2022)De~Santis, Vallet, and
  Gomes]{de_santis_environment_2021}
De~Santis,~M.; Vallet,~V.; Gomes,~A. S.~P. Environment {Effects} on {X}-{Ray}
  {Absorption} {Spectra} {With} {Quantum} {Embedded} {Real}-{Time}
  {Time}-{Dependent} {Density} {Functional} {Theory} {Approaches}. \emph{Front.
  Chem.} \textbf{2022}, \emph{10}, 823246, DOI:
  \doi{10.3389/fchem.2022.823246}\relax
\mciteBstWouldAddEndPuncttrue
\mciteSetBstMidEndSepPunct{\mcitedefaultmidpunct}
{\mcitedefaultendpunct}{\mcitedefaultseppunct}\relax
\EndOfBibitem
\bibitem[Koh(2020)]{koh_development_2020}
Koh,~K.~J. Development of {RT}-{TDDFT} for the {Interaction} {With} the
  {Explicit} {Solvent} and for {Correct} {Description} of {Excitation}
  {Process} - {ProQuest}. 2020;
  \url{https://www.proquest.com/docview/2624742534?pq-origsite=gscholar&fromopenview=true},
  Graduate Program in Chemistry and Biochemistry\relax
\mciteBstWouldAddEndPuncttrue
\mciteSetBstMidEndSepPunct{\mcitedefaultmidpunct}
{\mcitedefaultendpunct}{\mcitedefaultseppunct}\relax
\EndOfBibitem
\bibitem[Koh \latin{et~al.}(2017)Koh, Nguyen-Beck, and
  Parkhill]{koh_accelerating_2017}
Koh,~K.~J.; Nguyen-Beck,~T.~S.; Parkhill,~J. Accelerating {Realtime} {TDDFT}
  with {Block}-{Orthogonalized} {Manby}–{Miller} {Embedding} {Theory}.
  \emph{J. Chem. Theory Comput.} \textbf{2017}, \emph{13}, 4173--4178, DOI:
  \doi{10.1021/acs.jctc.7b00494}\relax
\mciteBstWouldAddEndPuncttrue
\mciteSetBstMidEndSepPunct{\mcitedefaultmidpunct}
{\mcitedefaultendpunct}{\mcitedefaultseppunct}\relax
\EndOfBibitem
\bibitem[Vignale(2008)]{vignale_real-time_2008}
Vignale,~G. Real-time resolution of the causality paradox of time-dependent
  density-functional theory. \emph{Phys. Rev. A} \textbf{2008}, \emph{77},
  062511, DOI: \doi{10.1103/PhysRevA.77.062511}\relax
\mciteBstWouldAddEndPuncttrue
\mciteSetBstMidEndSepPunct{\mcitedefaultmidpunct}
{\mcitedefaultendpunct}{\mcitedefaultseppunct}\relax
\EndOfBibitem
\bibitem[Maitra(2016)]{maitra_perspective_2016}
Maitra,~N.~T. Perspective: {Fundamental} aspects of time-dependent density
  functional theory. \emph{J. Chem. Phys.} \textbf{2016}, \emph{144}, 220901,
  DOI: \doi{10.1063/1.4953039}\relax
\mciteBstWouldAddEndPuncttrue
\mciteSetBstMidEndSepPunct{\mcitedefaultmidpunct}
{\mcitedefaultendpunct}{\mcitedefaultseppunct}\relax
\EndOfBibitem
\bibitem[Jakowski and Morokuma(2009)Jakowski, and
  Morokuma]{jakowski_liouvillevon_2009}
Jakowski,~J.; Morokuma,~K. Liouville–von {Neumann} molecular dynamics.
  \emph{J. Chem. Phys.} \textbf{2009}, \emph{130}, 224106, DOI:
  \doi{10.1063/1.3152120}\relax
\mciteBstWouldAddEndPuncttrue
\mciteSetBstMidEndSepPunct{\mcitedefaultmidpunct}
{\mcitedefaultendpunct}{\mcitedefaultseppunct}\relax
\EndOfBibitem
\bibitem[Li \latin{et~al.}(2005)Li, Smith, Markevitch, Romanov, Levis, and
  Schlegel]{li_time-dependent_2005}
Li,~X.; Smith,~S.~M.; Markevitch,~A.~N.; Romanov,~D.~A.; Levis,~R.~J.;
  Schlegel,~H.~B. A time-dependent {Hartree}–{Fock} approach for studying the
  electronic optical response of molecules in intense fields. \emph{Phys. Chem.
  Chem. Phys.} \textbf{2005}, \emph{7}, 233--239, DOI:
  \doi{10.1039/B415849K}\relax
\mciteBstWouldAddEndPuncttrue
\mciteSetBstMidEndSepPunct{\mcitedefaultmidpunct}
{\mcitedefaultendpunct}{\mcitedefaultseppunct}\relax
\EndOfBibitem
\bibitem[Turney \latin{et~al.}(2012)Turney, Simmonett, Parrish, Hohenstein,
  Evangelista, Fermann, Mintz, Burns, Wilke, Abrams, Russ, Leininger, Janssen,
  Seidl, Allen, Schaefer, King, Valeev, Sherrill, and
  Crawford]{turney_psi4_2012}
Turney,~J.~M.; Simmonett,~A.~C.; Parrish,~R.~M.; Hohenstein,~E.~G.;
  Evangelista,~F.~A.; Fermann,~J.~T.; Mintz,~B.~J.; Burns,~L.~A.; Wilke,~J.~J.;
  Abrams,~M.~L.; Russ,~N.~J.; Leininger,~M.~L.; Janssen,~C.~L.; Seidl,~E.~T.;
  Allen,~W.~D.; Schaefer,~H.~F.; King,~R.~A.; Valeev,~E.~F.; Sherrill,~C.~D.;
  Crawford,~T.~D. Psi4: an open-source \textit{ab initio} electronic structure
  program: {Psi4}: an electronic structure program. \emph{Wiley Interdiscip.
  Rev.: Comput. Mol. Sci.} \textbf{2012}, \emph{2}, 556--565, DOI:
  \doi{10.1002/wcms.93}\relax
\mciteBstWouldAddEndPuncttrue
\mciteSetBstMidEndSepPunct{\mcitedefaultmidpunct}
{\mcitedefaultendpunct}{\mcitedefaultseppunct}\relax
\EndOfBibitem
\bibitem[Smith \latin{et~al.}(2018)Smith, Burns, Sirianni, Nascimento, Kumar,
  James, Schriber, Zhang, Zhang, Abbott, Berquist, Lechner, Cunha, Heide,
  Waldrop, Takeshita, Alenaizan, Neuhauser, King, Simmonett, Turney, Schaefer,
  Evangelista, DePrince, Crawford, Patkowski, and Sherrill]{smith_p_2018}
Smith,~D. G.~A.; Burns,~L.~A.; Sirianni,~D.~A.; Nascimento,~D.~R.; Kumar,~A.;
  James,~A.~M.; Schriber,~J.~B.; Zhang,~T.; Zhang,~B.; Abbott,~A.~S.;
  Berquist,~E.~J.; Lechner,~M.~H.; Cunha,~L.~A.; Heide,~A.~G.; Waldrop,~J.~M.;
  Takeshita,~T.~Y.; Alenaizan,~A.; Neuhauser,~D.; King,~R.~A.;
  Simmonett,~A.~C.; Turney,~J.~M.; Schaefer,~H.~F.; Evangelista,~F.~A.;
  DePrince,~A.~E.; Crawford,~T.~D.; Patkowski,~K.; Sherrill,~C.~D. Psi4NumPy:
  {An} {Interactive} {Quantum} {Chemistry} {Programming} {Environment} for
  {Reference} {Implementations} and {Rapid} {Development}. \emph{J. Chem.
  Theory Comput.} \textbf{2018}, \emph{14}, 3504--3511, DOI:
  \doi{10.1021/acs.jctc.8b00286}\relax
\mciteBstWouldAddEndPuncttrue
\mciteSetBstMidEndSepPunct{\mcitedefaultmidpunct}
{\mcitedefaultendpunct}{\mcitedefaultseppunct}\relax
\EndOfBibitem
\bibitem[Press and Teukolsky(2007)Press, and Teukolsky]{press_numerical_2007}
Press,~W.~H.; Teukolsky,~S.~A. \emph{Numerical {Recipes} 3rd {Edition}: {The}
  {Art} of {Scientific} {Computing}}; Cambridge University Press, 2007;
  Google-Books-ID: 1aAOdzK3FegC\relax
\mciteBstWouldAddEndPuncttrue
\mciteSetBstMidEndSepPunct{\mcitedefaultmidpunct}
{\mcitedefaultendpunct}{\mcitedefaultseppunct}\relax
\EndOfBibitem
\bibitem[Meng and Kaxiras(2008)Meng, and Kaxiras]{meng_real-time_2008}
Meng,~S.; Kaxiras,~E. Real-time, local basis-set implementation of
  time-dependent density functional theory for excited state dynamics
  simulations. \emph{J. Chem. Phys.} \textbf{2008}, \emph{129}, 054110, DOI:
  \doi{10.1063/1.2960628}\relax
\mciteBstWouldAddEndPuncttrue
\mciteSetBstMidEndSepPunct{\mcitedefaultmidpunct}
{\mcitedefaultendpunct}{\mcitedefaultseppunct}\relax
\EndOfBibitem
\bibitem[Magnus(1954)]{magnus_exponential_1954}
Magnus,~W. On the exponential solution of differential equations for a linear
  operator. \emph{Commun. Pure Appl. Math.} \textbf{1954}, \emph{7}, 649--673,
  DOI: \doi{10.1002/cpa.3160070404}\relax
\mciteBstWouldAddEndPuncttrue
\mciteSetBstMidEndSepPunct{\mcitedefaultmidpunct}
{\mcitedefaultendpunct}{\mcitedefaultseppunct}\relax
\EndOfBibitem
\bibitem[Casas and Iserles(2006)Casas, and Iserles]{casas_explicit_2006}
Casas,~F.; Iserles,~A. Explicit {Magnus} expansions for nonlinear equations.
  \emph{J. Phys. A: Math. Gen.} \textbf{2006}, \emph{39}, 5445--5461, DOI:
  \doi{10.1088/0305-4470/39/19/S07}\relax
\mciteBstWouldAddEndPuncttrue
\mciteSetBstMidEndSepPunct{\mcitedefaultmidpunct}
{\mcitedefaultendpunct}{\mcitedefaultseppunct}\relax
\EndOfBibitem
\bibitem[Wesolowski and Warshel(1994)Wesolowski, and Warshel]{wesolowski1994ab}
Wesolowski,~T.; Warshel,~A. Ab initio free energy perturbation calculations of
  solvation free energy using the frozen density functional approach. \emph{J.
  Phys. Chem.} \textbf{1994}, \emph{98}, 5183--5187, DOI:
  \doi{https://doi.org/10.1021/j100071a003}\relax
\mciteBstWouldAddEndPuncttrue
\mciteSetBstMidEndSepPunct{\mcitedefaultmidpunct}
{\mcitedefaultendpunct}{\mcitedefaultseppunct}\relax
\EndOfBibitem
\bibitem[Thomas(1927)]{thomas_calculation_1927}
Thomas,~L.~H. The calculation of atomic fields. \emph{Mathematical Proceedings
  of the Cambridge Philosophical Society} \textbf{1927}, \emph{23}, 542--548,
  DOI: \doi{10.1017/S0305004100011683}\relax
\mciteBstWouldAddEndPuncttrue
\mciteSetBstMidEndSepPunct{\mcitedefaultmidpunct}
{\mcitedefaultendpunct}{\mcitedefaultseppunct}\relax
\EndOfBibitem
\bibitem[Laricchia \latin{et~al.}(2011)Laricchia, Fabiano, Constantin, and
  Della~Sala]{laricchia2011generalized}
Laricchia,~S.; Fabiano,~E.; Constantin,~L.; Della~Sala,~F. Generalized gradient
  approximations of the noninteracting kinetic energy from the semiclassical
  atom theory: Rationalization of the accuracy of the frozen density embedding
  theory for nonbonded interactions. \emph{J. Chem. Theory Comput.}
  \textbf{2011}, \emph{7}, 2439--2451, DOI:
  \doi{https://doi.org/10.1021/ct200382w}\relax
\mciteBstWouldAddEndPuncttrue
\mciteSetBstMidEndSepPunct{\mcitedefaultmidpunct}
{\mcitedefaultendpunct}{\mcitedefaultseppunct}\relax
\EndOfBibitem
\bibitem[Lembarki and Chermette(1994)Lembarki, and
  Chermette]{lembarki_obtaining_1994}
Lembarki,~A.; Chermette,~H. Obtaining a gradient-corrected kinetic-energy
  functional from the {Perdew}-{Wang} exchange functional. \emph{Phys. Rev. A}
  \textbf{1994}, \emph{50}, 5328--5331, DOI:
  \doi{10.1103/PhysRevA.50.5328}\relax
\mciteBstWouldAddEndPuncttrue
\mciteSetBstMidEndSepPunct{\mcitedefaultmidpunct}
{\mcitedefaultendpunct}{\mcitedefaultseppunct}\relax
\EndOfBibitem
\bibitem[Wesolowski and Weber(1996)Wesolowski, and Weber]{FaT}
Wesolowski,~T.~A.; Weber,~J. Kohn-Sham equations with constrained electron
  density: an iterative evaluation of the ground-state electron density of
  interacting molecules. \emph{Chem. Phys. Lett.} \textbf{1996}, \emph{248},
  71--76, DOI: \doi{https://doi.org/10.1016/0009-2614(95)01281-8}\relax
\mciteBstWouldAddEndPuncttrue
\mciteSetBstMidEndSepPunct{\mcitedefaultmidpunct}
{\mcitedefaultendpunct}{\mcitedefaultseppunct}\relax
\EndOfBibitem
\bibitem[Manby \latin{et~al.}(2012)Manby, Stella, Goodpaster, and
  Miller]{manby_simple_2012}
Manby,~F.~R.; Stella,~M.; Goodpaster,~J.~D.; Miller,~T.~F. A Simple, Exact
  Density Functional-Theory Embedding Scheme. \emph{J. Chem. Theory Comput.}
  \textbf{2012}, \emph{8}, 2564--2568, DOI: \doi{10.1021/ct300544e}\relax
\mciteBstWouldAddEndPuncttrue
\mciteSetBstMidEndSepPunct{\mcitedefaultmidpunct}
{\mcitedefaultendpunct}{\mcitedefaultseppunct}\relax
\EndOfBibitem
\bibitem[Pipolo \latin{et~al.}(2014)Pipolo, Corni, and
  Cammi]{pipolo_cavity_2014}
Pipolo,~S.; Corni,~S.; Cammi,~R. The cavity electromagnetic field within the
  polarizable continuum model of solvation. \emph{J. Chem. Phys.}
  \textbf{2014}, \emph{140}, 164114, DOI: \doi{10.1063/1.4871373}\relax
\mciteBstWouldAddEndPuncttrue
\mciteSetBstMidEndSepPunct{\mcitedefaultmidpunct}
{\mcitedefaultendpunct}{\mcitedefaultseppunct}\relax
\EndOfBibitem
\bibitem[Gomes \latin{et~al.}(2008)Gomes, Jacob, and Visscher]{Gomes2008}
Gomes,~A. S.~P.; Jacob,~C.~R.; Visscher,~L. Calculation of local excitations in
  large systems by embedding wave-function theory in density-functional theory.
  \emph{Phys. Chem. Chem. Phys.} \textbf{2008}, \emph{10}, 5353, DOI:
  \doi{10.1039/b805739g}\relax
\mciteBstWouldAddEndPuncttrue
\mciteSetBstMidEndSepPunct{\mcitedefaultmidpunct}
{\mcitedefaultendpunct}{\mcitedefaultseppunct}\relax
\EndOfBibitem
\bibitem[H\"{o}fener \latin{et~al.}(2012)H\"{o}fener, Severo Pereira~Gomes, and
  Visscher]{Hfener2012}
H\"{o}fener,~S.; Severo Pereira~Gomes,~A.; Visscher,~L. Molecular properties
  via a subsystem density functional theory formulation: A common framework for
  electronic embedding. \emph{J. Chem. Phys.} \textbf{2012}, \emph{136}, DOI:
  \doi{10.1063/1.3675845}\relax
\mciteBstWouldAddEndPuncttrue
\mciteSetBstMidEndSepPunct{\mcitedefaultmidpunct}
{\mcitedefaultendpunct}{\mcitedefaultseppunct}\relax
\EndOfBibitem
\bibitem[Konecny \latin{et~al.}(2016)Konecny, Kadek, Komorovsky, Malkina, Ruud,
  and Repisky]{konecny_acceleration_2016}
Konecny,~L.; Kadek,~M.; Komorovsky,~S.; Malkina,~O.~L.; Ruud,~K.; Repisky,~M.
  Acceleration of {Relativistic} {Electron} {Dynamics} by {Means} of {X2C}
  {Transformation}: {Application} to the {Calculation} of {Nonlinear} {Optical}
  {Properties}. \emph{J. Chem. Theory Comput.} \textbf{2016}, \emph{12},
  5823--5833, DOI: \doi{10.1021/acs.jctc.6b00740}\relax
\mciteBstWouldAddEndPuncttrue
\mciteSetBstMidEndSepPunct{\mcitedefaultmidpunct}
{\mcitedefaultendpunct}{\mcitedefaultseppunct}\relax
\EndOfBibitem
\bibitem[Repisky \latin{et~al.}(2015)Repisky, Konecny, Kadek, Komorovsky,
  Malkin, Malkin, and Ruud]{repisky_excitation_2015}
Repisky,~M.; Konecny,~L.; Kadek,~M.; Komorovsky,~S.; Malkin,~O.~L.;
  Malkin,~V.~G.; Ruud,~K. Excitation {Energies} from {Real}-{Time}
  {Propagation} of the {Four}-{Component} {Dirac}–{Kohn}–{Sham} {Equation}.
  \emph{J. Chem. Theory Comput.} \textbf{2015}, \emph{11}, 980--991, DOI:
  \doi{10.1021/ct501078d}\relax
\mciteBstWouldAddEndPuncttrue
\mciteSetBstMidEndSepPunct{\mcitedefaultmidpunct}
{\mcitedefaultendpunct}{\mcitedefaultseppunct}\relax
\EndOfBibitem
\bibitem[Imambi \latin{et~al.}(2021)Imambi, Prakash, and
  Kanagachidambaresan]{imambi_pytorch_2021}
Imambi,~S.; Prakash,~K.~B.; Kanagachidambaresan,~G.~R. In \emph{Programming
  with {TensorFlow}: {Solution} for {Edge} {Computing} {Applications}};
  Prakash,~K.~B., Kanagachidambaresan,~G.~R., Eds.; {EAI}/{Springer}
  {Innovations} in {Communication} and {Computing}; Springer International
  Publishing: Cham, 2021; pp 87--104, DOI:
  \doi{10.1007/978-3-030-57077-4_10}\relax
\mciteBstWouldAddEndPuncttrue
\mciteSetBstMidEndSepPunct{\mcitedefaultmidpunct}
{\mcitedefaultendpunct}{\mcitedefaultseppunct}\relax
\EndOfBibitem
\bibitem[{TensorFlow Developers}(2021)]{tensorflow_developers_tensorflow_2021}
{TensorFlow Developers}, {TensorFlow}. 2021;
  \url{https://zenodo.org/record/4758419}\relax
\mciteBstWouldAddEndPuncttrue
\mciteSetBstMidEndSepPunct{\mcitedefaultmidpunct}
{\mcitedefaultendpunct}{\mcitedefaultseppunct}\relax
\EndOfBibitem
\bibitem[Michelucci(2019)]{michelucci_tensorflow_2019}
Michelucci,~U. \emph{Advanced {Applied} {Deep} {Learning}}; Apress: Berkeley,
  CA, 2019; pp 27--77, DOI: \doi{10.1007/978-1-4842-4976-5_2}\relax
\mciteBstWouldAddEndPuncttrue
\mciteSetBstMidEndSepPunct{\mcitedefaultmidpunct}
{\mcitedefaultendpunct}{\mcitedefaultseppunct}\relax
\EndOfBibitem
\bibitem[Davis \latin{et~al.}(2021)Davis, \latin{et~al.}
  others]{davis2021snakeviz}
Davis,~M., \latin{et~al.}  SnakeViz, an in-browser Python profile viewer.
  2021\relax
\mciteBstWouldAddEndPuncttrue
\mciteSetBstMidEndSepPunct{\mcitedefaultmidpunct}
{\mcitedefaultendpunct}{\mcitedefaultseppunct}\relax
\EndOfBibitem
\bibitem[Storchi \latin{et~al.}(2020)Storchi, De~Santis, and
  Belpassi]{foster_bertha_2020}
Storchi,~L.; De~Santis,~M.; Belpassi,~L. In \emph{Advances in {Parallel}
  {Computing}}; Foster,~I., Joubert,~G.~R., Kučera,~L., Nagel,~W.~E.,
  Peters,~F., Eds.; IOS Press, 2020; DOI: \doi{10.3233/APC200060}\relax
\mciteBstWouldAddEndPuncttrue
\mciteSetBstMidEndSepPunct{\mcitedefaultmidpunct}
{\mcitedefaultendpunct}{\mcitedefaultseppunct}\relax
\EndOfBibitem
\bibitem[De~Santis \latin{et~al.}(2020)De~Santis, Storchi, Belpassi, Quiney,
  and Tarantelli]{de_santis_pyberthart_2020}
De~Santis,~M.; Storchi,~L.; Belpassi,~L.; Quiney,~H.~M.; Tarantelli,~F.
  PyBERTHART: A Relativistic Real Time Four Component TDDFT Implementation
  Using Prototyping Techniques Based on Python. \emph{J. Chem. Theory Comput.}
  \textbf{2020}, \emph{16}, 2410--2429, DOI:
  \doi{10.1021/acs.jctc.0c00053}\relax
\mciteBstWouldAddEndPuncttrue
\mciteSetBstMidEndSepPunct{\mcitedefaultmidpunct}
{\mcitedefaultendpunct}{\mcitedefaultseppunct}\relax
\EndOfBibitem
\bibitem[Belpassi \latin{et~al.}(2020)Belpassi, De~Santis, Quiney, Tarantelli,
  and Storchi]{belpassi_bertha_2020}
Belpassi,~L.; De~Santis,~M.; Quiney,~H.~M.; Tarantelli,~F.; Storchi,~L.
  {BERTHA}: {Implementation} of a four-component {Dirac}–{Kohn}–{Sham}
  relativistic framework. \emph{J. Chem. Phys.} \textbf{2020}, \emph{152},
  164118, DOI: \doi{10.1063/5.0002831}\relax
\mciteBstWouldAddEndPuncttrue
\mciteSetBstMidEndSepPunct{\mcitedefaultmidpunct}
{\mcitedefaultendpunct}{\mcitedefaultseppunct}\relax
\EndOfBibitem
\bibitem[De~Santis \latin{et~al.}(2013)De~Santis, Martinez, Vallet, Gomes,
  Sorbelli, Storchi, Belpassi, Jacob, and Tarantelli]{Pybertha_nt}
De~Santis,~M.; Martinez,~J.; Vallet,~V.; Gomes,~A. S.~P.; Sorbelli,~D.;
  Storchi,~L.; Belpassi,~L.; Jacob,~C.~R.; Tarantelli,~F. BOMME-RT.
  \url{https://github.com/RelMBdev/pybertha/tree/numericaltest}, 2013\relax
\mciteBstWouldAddEndPuncttrue
\mciteSetBstMidEndSepPunct{\mcitedefaultmidpunct}
{\mcitedefaultendpunct}{\mcitedefaultseppunct}\relax
\EndOfBibitem
\bibitem[Smith \latin{et~al.}(2020)Smith, Burns, Simmonett, Parrish, Schieber,
  Galvelis, Kraus, Kruse, Di~Remigio, Alenaizan, James, Lehtola, Misiewicz,
  Scheurer, Shaw, Schriber, Xie, Glick, Sirianni, O’Brien, Waldrop, Kumar,
  Hohenstein, Pritchard, Brooks, Schaefer, Sokolov, Patkowski, DePrince,
  Bozkaya, King, Evangelista, Turney, Crawford, and Sherrill]{smith_p_2020}
Smith,~D. G.~A.; Burns,~L.~A.; Simmonett,~A.~C.; Parrish,~R.~M.;
  Schieber,~M.~C.; Galvelis,~R.; Kraus,~P.; Kruse,~H.; Di~Remigio,~R.;
  Alenaizan,~A.; James,~A.~M.; Lehtola,~S.; Misiewicz,~J.~P.; Scheurer,~M.;
  Shaw,~R.~A.; Schriber,~J.~B.; Xie,~Y.; Glick,~Z.~L.; Sirianni,~D.~A.;
  O’Brien,~J.~S.; Waldrop,~J.~M.; Kumar,~A.; Hohenstein,~E.~G.;
  Pritchard,~B.~P.; Brooks,~B.~R.; Schaefer,~H.~F.; Sokolov,~A.~Y.;
  Patkowski,~K.; DePrince,~A.~E.; Bozkaya,~U.; King,~R.~A.; Evangelista,~F.~A.;
  Turney,~J.~M.; Crawford,~T.~D.; Sherrill,~C.~D. Psi4 1.4: Open source
  software for high throughput quantum chemistry. \emph{J. Chem. Phys.}
  \textbf{2020}, \emph{152}, 184108, DOI: \doi{10.1063/5.0006002}\relax
\mciteBstWouldAddEndPuncttrue
\mciteSetBstMidEndSepPunct{\mcitedefaultmidpunct}
{\mcitedefaultendpunct}{\mcitedefaultseppunct}\relax
\EndOfBibitem
\bibitem[De~Santis \latin{et~al.}(2013)De~Santis, Martinez, Vallet, Gomes,
  Sorbelli, Storchi, Belpassi, Jacob, and Tarantelli]{BOMME-rt}
De~Santis,~M.; Martinez,~J.; Vallet,~V.; Gomes,~A. S.~P.; Sorbelli,~D.;
  Storchi,~L.; Belpassi,~L.; Jacob,~C.~R.; Tarantelli,~F. BOMME-RT.
  \url{https://github.com/RelMBdev/bomme_rt/tree/fock_helper_restart},
  2013\relax
\mciteBstWouldAddEndPuncttrue
\mciteSetBstMidEndSepPunct{\mcitedefaultmidpunct}
{\mcitedefaultendpunct}{\mcitedefaultseppunct}\relax
\EndOfBibitem
\bibitem[De~Santis \latin{et~al.}(2023)De~Santis, Martinez, Vallet, Gomes,
  Sorbelli, Storchi, Belpassi, Jacob, and Tarantelli]{Psi4rt-FDE-pytorch}
De~Santis,~M.; Martinez,~J.; Vallet,~V.; Gomes,~A. S.~P.; Sorbelli,~D.;
  Storchi,~L.; Belpassi,~L.; Jacob,~C.~R.; Tarantelli,~F. Pybertha-Psi4emdrt.
  \url{https://github.com/RelMBdev/pybertha/tree/psi4rt-pytorch-mixed_basis},
  2023\relax
\mciteBstWouldAddEndPuncttrue
\mciteSetBstMidEndSepPunct{\mcitedefaultmidpunct}
{\mcitedefaultendpunct}{\mcitedefaultseppunct}\relax
\EndOfBibitem
\bibitem[De~Santis \latin{et~al.}(2023)De~Santis, Martinez, Vallet, Gomes,
  Sorbelli, Storchi, Belpassi, Jacob, and Tarantelli]{Psi4rt-FDE-tensorflow}
De~Santis,~M.; Martinez,~J.; Vallet,~V.; Gomes,~A. S.~P.; Sorbelli,~D.;
  Storchi,~L.; Belpassi,~L.; Jacob,~C.~R.; Tarantelli,~F. Pybertha-Psi4emdrt.
  \url{https://github.com/RelMBdev/pybertha/tree/psi4rt-tensorflow-mixed_basis},
  2023\relax
\mciteBstWouldAddEndPuncttrue
\mciteSetBstMidEndSepPunct{\mcitedefaultmidpunct}
{\mcitedefaultendpunct}{\mcitedefaultseppunct}\relax
\EndOfBibitem
\bibitem[Becke(1993)]{becke_density-functional_1993}
Becke,~A.~D. Density-functional thermochemistry. {III}. {The} role of exact
  exchange. \emph{J. Chem. Phys.} \textbf{1993}, \emph{98}, 5648--5652, DOI:
  \doi{10.1063/1.464913}\relax
\mciteBstWouldAddEndPuncttrue
\mciteSetBstMidEndSepPunct{\mcitedefaultmidpunct}
{\mcitedefaultendpunct}{\mcitedefaultseppunct}\relax
\EndOfBibitem
\bibitem[Papajak \latin{et~al.}(2011)Papajak, Zheng, Xu, Leverentz, and
  Truhlar]{papajak_perspectives_2011}
Papajak,~E.; Zheng,~J.; Xu,~X.; Leverentz,~H.~R.; Truhlar,~D.~G. Perspectives
  on {Basis} {Sets} {Beautiful}: {Seasonal} {Plantings} of {Diffuse} {Basis}
  {Functions}. \emph{J. Chem. Theory Comput.} \textbf{2011}, \emph{7},
  3027--3034, DOI: \doi{10.1021/ct200106a}\relax
\mciteBstWouldAddEndPuncttrue
\mciteSetBstMidEndSepPunct{\mcitedefaultmidpunct}
{\mcitedefaultendpunct}{\mcitedefaultseppunct}\relax
\EndOfBibitem
\bibitem[Dunning(1970)]{dunning_gaussian_1970}
Dunning,~T.~H. Gaussian {Basis} {Functions} for {Use} in {Molecular}
  {Calculations}. {I}. {Contraction} of (9s5p) {Atomic} {Basis} {Sets} for the
  {First}-{Row} {Atoms}. \emph{J. Chem. Phys.} \textbf{1970}, \emph{53},
  2823--2833, DOI: \doi{10.1063/1.1674408}\relax
\mciteBstWouldAddEndPuncttrue
\mciteSetBstMidEndSepPunct{\mcitedefaultmidpunct}
{\mcitedefaultendpunct}{\mcitedefaultseppunct}\relax
\EndOfBibitem
\bibitem[Jacob \latin{et~al.}(2011)Jacob, Beyhan, Bulo, Gomes, Götz, Kiewisch,
  Sikkema, and Visscher]{jacob_pyadf_2011}
Jacob,~C.~R.; Beyhan,~S.~M.; Bulo,~R.~E.; Gomes,~A. S.~P.; Götz,~A.~W.;
  Kiewisch,~K.; Sikkema,~J.; Visscher,~L. {PyADF} - {A} scripting framework for
  multiscale quantum chemistry. \emph{J. Comput. Chem.} \textbf{2011},
  \emph{32}, 2328--2338, DOI: \doi{10.1002/jcc.21810}\relax
\mciteBstWouldAddEndPuncttrue
\mciteSetBstMidEndSepPunct{\mcitedefaultmidpunct}
{\mcitedefaultendpunct}{\mcitedefaultseppunct}\relax
\EndOfBibitem
\bibitem[Te~Velde \latin{et~al.}(2001)Te~Velde, Bickelhaupt, Baerends,
  Fonseca~Guerra, Van~Gisbergen, Snijders, and
  Ziegler]{te_velde_chemistry_2001}
Te~Velde,~G.; Bickelhaupt,~F.~M.; Baerends,~E.~J.; Fonseca~Guerra,~C.;
  Van~Gisbergen,~S. J.~A.; Snijders,~J.~G.; Ziegler,~T. Chemistry with {ADF}.
  \emph{J. Comput. Chem.} \textbf{2001}, \emph{22}, 931--967, DOI:
  \doi{10.1002/jcc.1056}\relax
\mciteBstWouldAddEndPuncttrue
\mciteSetBstMidEndSepPunct{\mcitedefaultmidpunct}
{\mcitedefaultendpunct}{\mcitedefaultseppunct}\relax
\EndOfBibitem
\bibitem[Van~Lenthe and Baerends(2003)Van~Lenthe, and
  Baerends]{van_lenthe_optimized_2003}
Van~Lenthe,~E.; Baerends,~E.~J. Optimized {Slater}‐type basis sets for the
  elements 1–118. \emph{J. Comput. Chem.} \textbf{2003}, \emph{24},
  1142--1156, DOI: \doi{10.1002/jcc.10255}\relax
\mciteBstWouldAddEndPuncttrue
\mciteSetBstMidEndSepPunct{\mcitedefaultmidpunct}
{\mcitedefaultendpunct}{\mcitedefaultseppunct}\relax
\EndOfBibitem
\bibitem[Bruner \latin{et~al.}(2016)Bruner, LaMaster, and
  Lopata]{bruner_accelerated_2016}
Bruner,~A.; LaMaster,~D.; Lopata,~K. Accelerated {Broadband} {Spectra} {Using}
  {Transition} {Dipole} {Decomposition} and {Padé} {Approximants}. \emph{J.
  Chem. Theory Comput.} \textbf{2016}, \emph{12}, 3741--3750, DOI:
  \doi{10.1021/acs.jctc.6b00511}\relax
\mciteBstWouldAddEndPuncttrue
\mciteSetBstMidEndSepPunct{\mcitedefaultmidpunct}
{\mcitedefaultendpunct}{\mcitedefaultseppunct}\relax
\EndOfBibitem
\bibitem[Zuehlsdorff and Isborn(2019)Zuehlsdorff, and
  Isborn]{zuehlsdorff_modeling_2019}
Zuehlsdorff,~T.~J.; Isborn,~C.~M. Modeling absorption spectra of molecules in
  solution. \emph{Int. J. Quantum Chem.} \textbf{2019}, \emph{119}, DOI:
  \doi{10.1002/qua.25719}\relax
\mciteBstWouldAddEndPuncttrue
\mciteSetBstMidEndSepPunct{\mcitedefaultmidpunct}
{\mcitedefaultendpunct}{\mcitedefaultseppunct}\relax
\EndOfBibitem
\bibitem[Wu and Ellis(1995)Wu, and Ellis]{wu_x-ray_1995}
Wu,~Y.; Ellis,~D.~E. X-ray absorption near-edge spectra and electronic
  structure of rhodium compounds. \emph{J. Phys.: Condens. Matter}
  \textbf{1995}, \emph{7}, 3973--3989, DOI:
  \doi{10.1088/0953-8984/7/20/016}\relax
\mciteBstWouldAddEndPuncttrue
\mciteSetBstMidEndSepPunct{\mcitedefaultmidpunct}
{\mcitedefaultendpunct}{\mcitedefaultseppunct}\relax
\EndOfBibitem
\bibitem[Smith and Saykally(2017)Smith, and Saykally]{smith_soft_2017}
Smith,~J.~W.; Saykally,~R.~J. Soft {X}-ray {Absorption} {Spectroscopy} of
  {Liquids} and {Solutions}. \emph{Chem. Rev.} \textbf{2017}, \emph{117},
  13909--13934, DOI: \doi{10.1021/acs.chemrev.7b00213}\relax
\mciteBstWouldAddEndPuncttrue
\mciteSetBstMidEndSepPunct{\mcitedefaultmidpunct}
{\mcitedefaultendpunct}{\mcitedefaultseppunct}\relax
\EndOfBibitem
\bibitem[Gaiduk and Galli(2017)Gaiduk, and Galli]{gaiduk_local_2017}
Gaiduk,~A.~P.; Galli,~G. Local and {Global} {Effects} of {Dissolved} {Sodium}
  {Chloride} on the {Structure} of {Water}. \emph{J. Phys. Chem. Lett.}
  \textbf{2017}, \emph{8}, 1496--1502, DOI:
  \doi{10.1021/acs.jpclett.7b00239}\relax
\mciteBstWouldAddEndPuncttrue
\mciteSetBstMidEndSepPunct{\mcitedefaultmidpunct}
{\mcitedefaultendpunct}{\mcitedefaultseppunct}\relax
\EndOfBibitem
\bibitem[Penfold \latin{et~al.}()Penfold, Curchod, Tavernelli, Abela,
  Rothlisberger, and Chergui]{Penfold_200x}
Penfold,~T.~J.; Curchod,~B.~F.; Tavernelli,~I.; Abela,~R.; Rothlisberger,~U.;
  Chergui,~M. Theoretical considerations for the simulation of X-ray absorption
  spectra: The effect of the solvent for a diplatinum complex.
  \url{https://drive.google.com/file/d/182vqPmPJ0f2jYcODWBsv-oHMpwXOP5Ln/view}\relax
\mciteBstWouldAddEndPuncttrue
\mciteSetBstMidEndSepPunct{\mcitedefaultmidpunct}
{\mcitedefaultendpunct}{\mcitedefaultseppunct}\relax
\EndOfBibitem
\end{mcitethebibliography}


\providecommand{\latin}[1]{#1}
\makeatletter
\providecommand{\doi}
  {\begingroup\let\do\@makeother\dospecials
  \catcode`\{=1 \catcode`\}=2 \doi@aux}
\providecommand{\doi@aux}[1]{\endgroup\texttt{#1}}
\makeatother
\providecommand*\mcitethebibliography{\thebibliography}
\csname @ifundefined\endcsname{endmcitethebibliography}
  {\let\endmcitethebibliography\endthebibliography}{}
\begin{mcitethebibliography}{2}
\providecommand*\natexlab[1]{#1}
\providecommand*\mciteSetBstSublistMode[1]{}
\providecommand*\mciteSetBstMaxWidthForm[2]{}
\providecommand*\mciteBstWouldAddEndPuncttrue
  {\def\EndOfBibitem{\unskip.}}
\providecommand*\mciteBstWouldAddEndPunctfalse
  {\let\EndOfBibitem\relax}
\providecommand*\mciteSetBstMidEndSepPunct[3]{}
\providecommand*\mciteSetBstSublistLabelBeginEnd[3]{}
\providecommand*\EndOfBibitem{}
\mciteSetBstSublistMode{f}
\mciteSetBstMaxWidthForm{subitem}{(\alph{mcitesubitemcount})}
\mciteSetBstSublistLabelBeginEnd
  {\mcitemaxwidthsubitemform\space}
  {\relax}
  {\relax}

\bibitem[De~Santis \latin{et~al.}(2020)De~Santis, Belpassi, Jacob, Severo
  Pereira~Gomes, Tarantelli, Visscher, and
  Storchi]{de_santis_environmental_2020}
De~Santis,~M.; Belpassi,~L.; Jacob,~C.~R.; Severo Pereira~Gomes,~A.;
  Tarantelli,~F.; Visscher,~L.; Storchi,~L. Environmental {Effects} with
  {Frozen}-{Density} {Embedding} in {Real}-{Time} {Time}-{Dependent} {Density}
  {Functional} {Theory} {Using} {Localized} {Basis} {Functions}. \emph{J. Chem.
  Theory Comput.} \textbf{2020}, \emph{16}, 5695--5711, DOI:
  \doi{10.1021/acs.jctc.0c00603}\relax
\mciteBstWouldAddEndPuncttrue
\mciteSetBstMidEndSepPunct{\mcitedefaultmidpunct}
{\mcitedefaultendpunct}{\mcitedefaultseppunct}\relax
\EndOfBibitem
\end{mcitethebibliography}

\end{document}